\documentstyle[graphicx]{mn}
\newcommand{\ls}
 {\mathrel{\hbox{\rlap{\hbox{\lower4pt\hbox{$\sim$}}}\hbox{$<$}}}}
\newcommand{\gs}
 {\mathrel{\hbox{\rlap{\hbox{\lower4pt\hbox{$\sim$}}}\hbox{$>$}}}}
\newcommand{\degg}{\hbox{$^\circ$}}

\newcommand{\et}{{\it et al.}\ }
\newcommand{\rosat}{{\it ROSAT}}

\newcommand{\asca}{{\it ASCA}}
\newcommand{\xte}{{\it RXTE}}

\newcommand{\nh}{$N_{\rm H}$}
\newcommand{\LX}{L$_{X}$}
\newcommand{\RL}{R$_{L}$}

\title[X-ray Spectra of a large sample of Quasars with \asca]
	{ X-ray Spectra of a large sample of Quasars with ASCA }
\author[J. Reeves \& M. Turner]
        {J.N.\ Reeves$^{1}$ and M.J.L.\ Turner$^{1}$\\
$^1$X-Ray Astronomy Group; Department of Physics and Astronomy;
Leicester University; Leicester LE1 7RH; U.K.\\
}

\date{Accepted March 2000}

\pagerange{\pageref{firstpage}--\pageref{lastpage}}
\pubyear{1999}

\begin{document}

\maketitle

\label{firstpage}

\begin{abstract}

The results from an X-ray spectral analysis of a large sample of quasars,
observed with ASCA, are presented. The sample was selected to include
all ASCA observations of  quasars, with z\ $>0.05$ and M$_{V}<-23.0$,
available up to January 1998. The data reduction
leaves 62 quasars, 35 radio-loud and 27 radio-quiet,
suitable for spectral analysis. Differences are found between 
the radio-quiet quasars (RQQs) and the radio-loud quasars (RLQs);
the RLQs have flatter X-ray spectra ($\Gamma\sim1.6$), with little iron
line emission or reflection and are more X-ray luminous than the softer
($\Gamma\sim1.9$) RQQs, in agreement with previous studies. 
A correlation between $\Gamma$ and optical H$\beta$ was also found for the
{\it radio-quiet} quasars 
in this sample, whereby the steepest X-ray spectra tend
to be found in those objects with narrow H$\beta$ widths. The
correlation is significant at $>$99\% confidence, confirming the
well-known trend between $\Gamma$ and H$\beta$ FWHM in Seyfert 1s (Brandt \et
1997), but at higher luminosities. 

Other spectral complexities are observed from this sample. A soft X-ray
excess, with blackbody temperatures in the range 100 - 300 eV, is seen in
many low z radio-quiet quasars. In most cases the
temperatures are probably too hot to originate directly from the disk and
could imply that some reprocessing is involved.
Iron K line emission features are also found in
the RQQs; but often from partially ionised material.
Indeed in the highest luminosity RQQs there is neither evidence
for iron line emission nor the reflection component expected from
disk reflection models. These observations can be explained by an
increase in the quasar accretion rate with luminosity, leading to an
increase in
the ionisation state of the surface layers of the disk. 
The occurrence of ionised or `warm' absorbers is rare in this sample, with
only 5 detections in low z objects. 
However excess
neutral X-ray absorption is found towards several of the high z,
predominantly
radio-loud, quasars. 
Although found to increase with quasar redshift, this `intrinsic'
absorption may be associated with radio-loud AGN.


\end{abstract}


\section{Introduction}

Discovered in 1963 (Schmidt 1963), quasars are the most luminous 
continuously emitting objects in the Universe and 
represent the high luminosity end of the class of objects known as
Active Galactic Nuclei (AGN). Like their lower luminosity cousins - Seyfert
1 galaxies - the bulk of the energy produced in quasars is thought to
arise from accretion onto a compact object (the putative
super-massive black hole). This central engine is also thought to be
where the X-rays, that are observed from both quasars and Seyfert 1s,
originate from. 

In one model, the UV photons produced by viscous
dissipation in an accretion disk are Comptonised to X-ray energies by
a hot corona above the surface of this disk (Haardt \& Maraschi
1993). These hard X-rays in turn
illuminate the accretion disk, being either `reflected' back towards the
observer or thermalised into optical or UV photons. Evidence for
these `reflection' features (in the form of an iron K$\alpha$ emission
line, Fe K absorption edge and
Compton scattering hump) is commonly observed in the X-ray spectral 
band in Seyfert 1 galaxies (e.g. Pounds \et 1990, Nandra \& Pounds
1994). The detection of these reflection features however, in quasars,
remains more ellusive
(Reeves \et 1997, Lawson \& Turner 1997). In the {\it radio-loud}
quasars the situation is somewhat further complicated by the presence
of a powerful relativistic radio-jet. In the X-ray band, these
radio-loud quasars have flatter X-ray spectral emission (e.g. Wilkes \&
Elvis
1987, Lawson \et 1992), and are generally more luminous than the
radio-quiet quasars. The radio-loud quasars also have little 
or no X-ray (iron) line emission; this is often interpreted in terms of the
Doppler boosting of the X-ray continuum, by the relativistic
jet (see Reeves \et 1997 and references therein).

This paper presents the results of a detailed spectral analysis of 68
quasars obtained from the \asca\ public archive. The aims are to
extend the results that were presented in Reeves \et (1997), which
contained a smaller sample of 24 objects. The objects
considered in this sample contain a roughly equal mix of both
radio-loud and radio-quiet quasars. The bigger sample, for instance, 
allows us to perform an investigation of the properties of 
iron K lines and reflection in quasars, which will be limited in some of
the objects by signal-to-noise. It also permits us to investigate the
properties of quasars over a large range of luminosity and redshift. A
further aim of the paper 
is to make the results of this analysis available to
the general community; the paper presents results 
from a considerable number of quasars that are currently unpublished. 

In the following section, the selection and properties of the
sample are discussed. Section 3 then outlines the spectral fitting that
was performed on the quasars in the sample. The following sections (4-7)
then present and discuss the results in terms of the X-ray continuum
emission, soft X-ray excesses from quasars, the properties of the iron
line and reflection
associated with the putative accretion disk and also the effects of
absorbing material on the quasar spectra. 
Values of $ H_0 = 50 $~km~s$^{-1}$~Mpc$^{-1}$ and $ q_0 = 0.5 $
are assumed and all fit parameters are given in the quasar rest frame.

\section{The ASCA sample of Quasars}

Data have been selected mainly from the \asca\ public archive, 
using all quasars that were available up until January
1998. The observations included in this paper are shown in table 1; in
general when there has been a multiple observation of a given quasar
we have taken the first observation. In total 68
quasars have been included in the sample, with the objects selected
predominantly type I AGN. This covers a wide range of redshift
(from z=0.06 to z=4.3), and also a wide range of luminosities
(M$_{V} = -23$ to -30 
and L$_{2-10keV}\sim10^{43}$ erg~s$^{-1}$ to $>$10$^{47}$
erg~s$^{-1}$). Note that redshift and luminosity cut-offs of z\ $>0.05$
and M$_{V}<-23.0$ respectively have been used to define the sample. The
distribution of quasar redshift for this sample is included as figure 1.
Note that at z\ $>3$, the sample is predominantly
made up of bright core-dominated {\it radio-loud} quasars. 

\begin{figure}
\begin{center}
\rotatebox{-90}{\includegraphics[width=5.5cm]{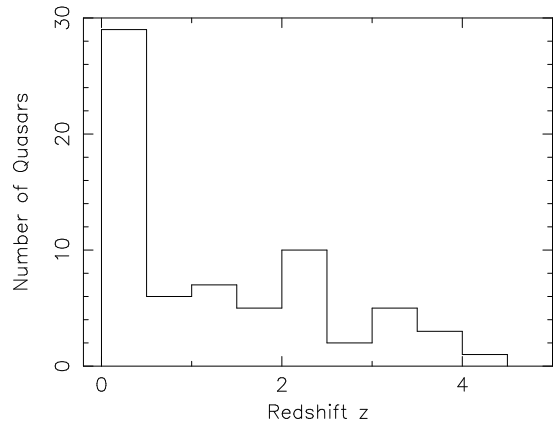}}
\end{center}
  \caption{Redshift distribution of quasars from the sample.}
\end{figure}

Standard (to conservative) \asca\ screening
criteria and data selection have been used in analysing the data, this
leaves 62 quasars that have sufficient
signal-to-noise for further spectral analysis and interpretation. The
selection criteria used in reducing these observations are outlined in
Reeves \et (1997). Of these 62
quasars, 35 are radio-loud and 27 are radio-quiet; according to the
definition of radio-loudness (R$_{L}$) used by Wilkes \& Elvis
(1987), where R$_{L}$ = log (F$_{5GHz}$/F$_{B}$). 
The cut-off between radio-quiet and radio-loud is
defined arbitrarily at R=1. The roughly even numbers of objects
in both classes allows us to investigate the properties within both.

Some objects, as described above, have been left
out of the subsequent analysis. 
The radio-quiet object IRAS P09104+4109 may be a buried
quasar, i.e. surrounded by a very high column of material; however most of
the X-ray emission is thought to originate from a surrounding cluster and
X-ray luminous cooling flow (Fabian \& Crawford 1995). 
Similarly the RQQ E 0015+162 is strongly contaminated by the
powerful X-ray emitting cluster CL 0016+16 (e.g. Neumann \& Bohringer
1997) and the spatial resolution of ASCA is not sufficient to resolve these two
objects. Therefore further analysis of these 2 sources have been
excluded from our sample. Finally spectra have only been
extracted where the source is detected at the 5$\sigma$ level in the ASCA
detectors. This excludes the radio-quiet quasars QSO 0215-504, MS
12487+5706 and QSO 1725+503 from further study. The radio-quiet broad
absorption line (BAL) quasar
PHL 5200 has only been detected in the GIS3 and SIS0 detectors, but not
GIS2
nor SIS1. As the data from this quasar contain few counts with which to
constrain the spectral form, this object has also been excluded from
further
study in the sample. 

\section{X-ray Spectral Fitting}

The \asca\ data
reduction process provides four seperate spectra, one for each of the
four (GIS and SIS) instruments. 
Spectral fitting was performed by simultaneously fitting the
data from each instrument, allowing the relative normalisations
to vary as necessary. The processed spectra were fitted with a
spectral model using the X-ray spectral fitting software XSPEC \textsc
{v10.0}, over the energy range
0.8-10 keV for the GIS instruments and from 0.6-10 keV for the SIS. 

The standard model used consists of a
power-law with neutral absorption, together with a Gaussian-shaped
narrow ($\sigma=0.01$~keV) Fe K line at
6.4~keV in the rest frame of the quasar. The absorption column 
that was fitted has
two components; (i) the Galactic column density which was fixed in value in
the z=0 frame, and (ii) the intrinsic absorption of the quasar in its rest
frame, which was allowed to vary in value. The Galactic column
densities have been obtained from the \textsc{getnh} program, run
under the \textsc{xanadu}
system, which uses data from Stark \et (1992) and other surveys;
absorption cross-sections have been taken from Morrison \& McCammon (1983).
Where possible more accurate values of \nh\ towards a particular quasar
have
been taken from individual observations using either Elvis \et (1989) or
Murphy \et (1996), which use higher spatial resolution.

\begin{table*}
\centering
\caption{ List of Observations. 
RLQ: Radio-Loud Quasar, RQQ: Radio-Quiet Quasar. 
{\it a} Galactic Column Density in units 10$^{20}$\,cm$^{-2}$.}

\begin{tabular}{lllrrlllllll}\hline
Source & Common & Object & \multicolumn{2}{c} {Co-ordinates (2000)} &
Obs Date & Duration & z & M$_{V}$ & R$_{L}$ & {\it N}$_{\rm H}^a$ \\
& Name & Class & RA & Dec & (yy.ddd) & (ksec) \\\hline
 
0014+813 & S5 & RLQ & 00 17 08.0 & 81 35 07 & 93.302 & 68.1 & 3.41 &
-29.7 & 2.33 & 14.6 \\ 
0015+162 & E & RQQ & 00 18 31.8 & 16 29 26 & 93.200 & 91.5 & 0.554 &
-24.0 & 0.71 & 4.22 \\
0050+124 & IZWI & RQQ & 00 53 34.8 & 12 41 36 & 95.196 & 93.6 & 0.061
& -23.4 & -0.59 & 5.07 \\
0054+144 & PHL 909 & RQQ & 00 57 09.8 & 14 46 11 & 96.018 & 45.8 &
0.17 & -23.1 & -0.02 & 4.2 \\
0057-222 & TON S180 & RQQ & 00 57 20.1 & -22 22 55 & 96.192 & 96.5 &
0.062 & -23.0 & $<$-0.70 &  1.57 \\
0205+024 & NAB & RQQ & 02 07 49.8 & 02 42 55 & 96.018 & 118.6 &  0.155
& -24.2 & -0.21 & 2.99 \\
0208-512 & PKS & RLQ & 02 10 46.2 & 51 01 02 & 95.015 & 32.2 & 1.003 &
-26.3 & 3.63 & 2.49 \\
0215-504 & QSO & RQQ & 02 17 25.3 & -50 15 32 & 95.192 & 100.8 & 2.620
& -25.7 & $<$1.17 & 2.47 & \ \\
0232-04 & PKS/PHL1377 & RLQ & 02 35 07.2 & -04 02 05 & 94.220 & 104.5
& 1.438 & -27.5 & 2.75 & 3.28 \\
0237-233 & PKS & RLQ & 02 40 08.1 & -23 09 17 & 94.222 & 104.4 & 2.224
& -28.0 & 3.50 & 2.29 \\
0332-403 & PKS & RLQ & 03 34 13.7 & -40 08 26 & 94.224 & 45.1 & 1.445
& -25.4 & 4.17 & 2.18 \\
0333+321 & NRAO140 & RLQ & 03 36 30.1 & 32 18 28 & 94.032 & 92.6 &
1.258 & -27.3 & 2.85 & 14.2 \\ 
0348-120 & PKS/H & RLQ & 03 51 10.8 & -11 53 20 & 96.036 & 109.0 & 1.520 &
-25.1 & 3.56 & 4.09 \\
0410+110 & 3C 109.0 & RLQ & 04 13 40.3 & 11 12 14 & 95.240 & 103.9 &
0.306 & -23.9 & 3.25 & 30.0 \\
0420-01 & PKS & RLQ & 04 23 15.7 & -01 20 33 & 97.243 & 60.9 & 0.915 &
-25.6 & 3.52 & 7.47 \\
0438-436 & PKS & RLQ & 04 40 17.1 & -43 33 09 & 93.194 & 85.8 & 2.852
& -26.3 & 4.75 & 1.3 \\
0440-285 & PKS & RLQ & 04 42 37.5 & -28 25 29 & 97.268 & 57.4 & 1.952
& -25.5 & 3.48 & 2.93 \\
0443-282 & X & RLQ & 04 43 19.3 & -28 20 04 & 97.268 & 57.4 & 0.156 &
-24.0 & 1.33 & 2.93 & \ \\ 
0449-183 & E & RQQ & 04 51 38.8 & -18 18 55 & 94.063 & 80.3 & 0.338 &
-23.0 & 0.97 & 3.77 \\
0528+134 & PKS & RLQ & 05 30 56.4 & 13 31 54 & 95.066 & 62.9 & 2.07 &
-26.9 & 4.03 & 23 \\
0537-286 & PKS & RLQ & 05 39 54.2 & -28 39 55 & 94.074 & 102.7 & 3.11
& -25.2 & 4.35 & 1.95 \\
0558-504 & PKS & RLQ & 05 59 47.4 & -50 26 51 & 96.249 & 97.6 & 0.137
& -24.4 & 1.34 & 5.28 \\
0836+715 & S5 & RLQ & 08 41 24.4 & 70 53 41 & 95.076 & 40.1 & 2.17 &
-28.2 & 3.33 & 2.83 \\
0903+169 & 3C 215.0 & RLQ & 09 06 31.8 & 16 46 11 & 95.308 & 84.8 &
0.411 & -23.4 & 3.23 & 3.37 \\
0910+410 & IRAS & RQQ & 09 13 45.5 & 40 56 26 & 93.316 & 91.8 & 0.442
& -23.8 & 0.80 & 1.82 \\
1029-1401 & HE & RQQ & 10 31 54.3 & -14 16 52 & 95.342 & 107.0 & 0.086
& -24.2 & -0.12 & 5.27 \\
1101-264 & Q & RQQ & 11 03 25.2 & -26 45 15 & 96.168 & 42.8 & 2.152 &
-28.7 & 0.61 & 6.38 \\
1104-1805 & HE & RQQ & 11 06 33.5 & -18 21 24 & 96.162 & 107.8 & 2.319
& -28.7 & $<$-0.96 & 4.24 \\
1114+445 & PG & RQQ & 11 17 06.2 & 44 13 33 & 96.126 & 172.2 & 0.144 &
-23.3 & -0.88 & 1.93 \\
1116+215 & PG & RQQ & 11 19 08.7 & 21 19 17 & 95.139 & 43.1 & 0.177 &
-24.6 & -0.28 & 2.2 \\
1148+549 & PG & RQQ & 11 51 20.4 & 54 37 32 & 95.341 & 81.6 & 0.968 &
-27.4 & -0.15 & 1.17 \\
1211+143 & PG & RQQ & 12 14 17.6 & 14 03 12 & 93.154 & 91.8 & 0.085 &
-23.8 & -0.92 & 2.83 \\
1216+069 & PG & RQQ & 12 19 20.6 & 06 38 37 & 95.359 & 45.5 & 0.334 &
-25.4 & 0.41 & 1.91 \\ 
1219+755 & Mrk205 & RQQ & 12 21 44.0 & 75 18 37 & 94.337 & 75.1 &
0.070 & -23.4 & -0.27 & 2.74 \\
1226+023 & 3C 273 & RLQ & 12 29 06.6 & 02 03 08 & 93.361 & 29.2 &
0.158 & -26.7 & 2.92 & 1.9 \\
1247+268 & PG & RQQ & 12 50 05.6 & 26 31 06 & 95.168 & 100.8 & 2.041 &
-28.7 & -0.44 & 0.86 \\
12487+5706 & MS & RQQ & 12 50 58.0 & 56 49 55 & 94.143 & 86.2 & 1.843
& -25.3 & $<$1.07 & 1.22 \\
1253-055 & 3C 279 & RLQ & 12 56 11.1 & -05 47 21 & 93.172 & 92.2 &
0.538 & -24.3 & 4.60 & 2.22 \\
1308+326 & B2 & RLQ & 13 10 28.6 & 32 20 43 & 96.162 & 86.8 & 0.997 &
-28.0 & 2.65 & 1.10 \\
13349+2438 & IRAS & RQQ & 13 37 18.7 & 24 23 03 & 95.178 & 49.9 & 0.107
& -23.7 & 0.20 & 1.08 \\
13434+0001 & RD J & RQQ & 13 43 23.6 & 00 00 55 & 96.012 & 252.4 &
2.350 & -23.8 & $<$2.27 & 3.26 \\
1404+226 & PG & RQQ & 14 06 21.8 & 22 23 46 & 94.194 & 100.6 & 0.098 &
-23.1 & -0.34 & 2.0 \\
1407+265 & PG & RQQ & 14 09 23.7 & 26 28 20 & 93.183 & 84.8 & 0.944 &
-27.5 & 0.45 & 1.4 \\ 
1416-129 & PG & RQQ & 14 19 03.7 & -13 10 44 & 94.210 & 79.1 & 0.129 &
-24.3 & -0.46 & 7.2 \\
1422+231 & B & RLQ & 14 24 38.1 & 22 56 00 & 95.014 & 86.5 & 3.62 &
-29.0 & 2.65 & 2.88 \\
1425+267 & PG & RLQ & 14 27 35.6 & 26 32 13 & 96.208 & 91.8 & 0.366 &
-25.6 & 1.75 & 1.54 \\ 
1440+356 & Mrk478 & RQQ & 14 42 07.4 & 35 26 22 & 95.183 & 86.8 &
0.079 & -23.1 & -0.01 & 1.24 \\
1508+571 & Q & RLQ & 15 10 02.1 & 57 03 04 & 95.061 & 165.8 & 4.3 &
-26.8 & 3.36 & 1.34 \\
1510-089 & PKS & RLQ & 15 12 50.4 & -09 06 00 & 96.233 & 64.4 & 0.360
& -25.1 & 3.47 & 8.38 \\
1559+089 & Q & RQQ & 16 02 22.5 & 08 45 36 & 96.236 & 63.7 & 2.269 &
-26.2 & $<$0.87 & 4.22 \\
1614+051 & PKS & RLQ & 16 16 37.6 & 04 59 32 & 94.214 & 92.4 & 3.21 &
-25.8 & 4.12 & 5.21 \\ 
1633+38 & 4C38.41/B2 & RLQ & 16 35 15.3 & 38 08 03 & 96.064 & 22.2 &
1.814 & -26.3 & 4.16 & 1.02 \\ 
1634+706 & PG & RQQ & 16 34 25.0 & 70 35 27 & 94.122 & 63.1 & 1.334 &
-28.9 & -0.59 & 5.74 \\
1718+481 & PG & RLQ & 17 19 38.6 & 48 08 07 & 94.121 & 63.5 & 1.084 &
-28.1 & 1.56 & 2.61 \\
1725+503 & QSO & RQQ & 17 26 57.5 & 50 15 48 & 95.063 & 85.6 & 2.100 &
-24.2 & $<$1.59 & 3.24 \\
17254-1413 & PDS 456 & RQQ & 17 28 19.8 & -14 15 56 & 98.066 & 114.2
& 0.184 & -27.1 & -0.46 & 24.0 \\
1745+624 & Q & RLQ & 17 46 14.7 & 62 27 04 & 94.239 & 55.5 & 3.87 &
-26.2 & $<$1.59 & 3.51 \\
1821+643 & E & RQQ & 18 21 59.2 & 64 21 07 & 93.170 & 91.5 & 0.297 &
-26.8 & 0.013 & 5.6 \\
1946+765 & HS & RQQ & 19 44 54.8 & 77 05 52 & 95.294 & 78.5 & 3.02 &
-29.7 & -0.51 & 7.4 \\
2000-330 & PKS & RLQ & 20 03 23.9 & -32 51 47 & 95.286 & 67.9 & 3.777
& -26.5 & 4.05 & 7.8 \\
2019+098 & 3C 411.0 & RLQ & 20 22 08.3 & 10 01 12 & 94.296 & 56.8 &
0.469 & -23.1 & 3.85 & 7.92 \\\hline

\end{tabular}
\end{table*}
\newpage
\begin{table*}
\begin{tabular}{lllrrlllllll}
Table 1 continued \\
\hline
Source & Common & Object & \multicolumn{2}{c} {Co-ordinates (2000)} &
Obs Date & Drn & z & M$_{V}$ & R$_{L}$ & {\it N}$_{\rm H}^a$ \\
& Name & Class & RA & Dec & (yy.ddd) & (ksec) \\\hline 

2043+749 & 4C 74.26/S5 & RLQ & 20 42 37.1 & 75 08 01 & 94.296 & 43.7 &
0.104 & -24.3 & 1.60 & 13.3 \\
2126-158 & PKS & RLQ & 21 29 12.0 & -15 38 41 & 93.136 & 105.5 & 3.27
& -28.0 & 3.37 & 4.85 \\
2149-306 & PKS & RLQ & 21 51 55.3 & -30 27 53 & 94.299 & 46.5 & 2.345
& -26.3 & 3.83 & 1.93 \\
2225-055 & PHL 5200 & RQQ & 22 28 30.3 & -05 18 55 & 94.172 & 55.0 &
1.98 & -27.0 & $<$0.03 & 4.07 \\
2230+114 & CTA 102 & RLQ & 22 32 36.3 & 11 43 51 & 95.340 & 45.0 &
1.037 & -26.1 & 3.78 & 5.05 \\
2251+113 & PKS/PG & RLQ & 22 54 10.4 & 11 36 38 & 96.153 & 73.4 &
0.323 & -25.3 & 2.43 & 5.53 \\ 
2251-178 & MR & RQQ & 22 54 05.7 & -17 34 54 & 93.310 & 21.6 & 0.068 &
-23.4 & -0.43 & 2.84 \\\hline

\end{tabular}
\end{table*}

\begin{table*}
\caption{
{\it a} units Flux 10$^{-12}$\,erg\,cm$^{-2}$\,s$^{-1}$ or
10$^{-15}$\,Wm$^{-2}$. Range of flux from 0.5-10keV. 
Luminosity units 10$^{45}$\,erg/s or 10$^{38}$\,W,
taken over the range from 2-10 keV in the quasar rest-frame.
{\it b} Intrinsic Column density, fitted in the quasar frame, units
10$^{21}$\,cm$^{-2}$\,. 
{\it c} Improvement in chi-squared upon allowing the column density to
become a free parameter in the spectral fit.
{\it d} Properties of the best fit Fe K$\alpha$ line. Units line flux
10$^{-14}$\,erg\,cm$^{-2}$\,s$^{-1}$\,. EQW is in eV, measured in the rest
frame of the quasar.
{\it e} Improvement in chi-squared upon adding the iron line (with narrow
width) to the spectral fit. Where significant the iron line energy is also
a free parameter (see
tables 5 for more detailed spectral fits.)
{\it f} Value of $\chi$$^{2}$\,/$\nu$\,, for the best fitting
spectrum.
{\it g} Photon index quoted over 2-10 keV band.
{\it h} Complex X-ray spectrum. Fitted over the 0.6-10 keV band, but $\Gamma$
does not including effects due to reprocessing features (see Reeves \et
2000 for further details).}

\begin{tabular}{llccccccccc}\hline
Quasar & Radio & $\Gamma$ & \multicolumn{2}{c} {Continuum$^a$} &
{\it N}$_{\rm H}^b$ & $\Delta\chi^{2}$$_{c}$ & Line Flux$^d$ & EQW &
$\Delta\chi^{2}$$_{e}$ &
$\chi^2/\nu^f$ \\
Name & Class & \ & Flux & L$_{2-10}$ \\\hline

S5~0014+813 & RL & 1.69$\pm$0.05 & 3.80 & 107.6 &
64.5$^{+27.4}_{-24.5}$ & 19.8 & $<1.3$ & $<93$ & 0.0 & 329.8/319 \\

E~0015+162 & RQ & 1.99$\pm$0.09 & 2.92 & 2.81 &
6.2$^{+2.5}_{-1.9}$ & 9.3 & $<3.36$ & $<174$ & 0.3 & 267.3/258 \\

IZWI & RQ & 2.37$\pm$0.05 & 6.81 & 0.055 &
0.31$^{+0.16}_{-0.16}$ & 8.9 & 12.0 & 483$^{+212}_{-211}$ & 11.9 &
604.3/608 \\

PHL909 & RQ & 1.11$\pm$0.11 & 4.1 & 0.39 & $<0.35$ & 0.0 &
5.25 & 158$^{+94}_{-94}$ & 4.5 & 257.4/247 \\

TON\ S180$^g$ & RQ & 2.31$\pm$0.07 & 16.1 & 0.088 &
$<0.20$ & 0.0 & 5.0 & 123$^{+75}_{-75}$ & 8.2 & 978.7/977 \\

NAB~0205+024 & RQ & 2.10$\pm$0.08 & 5.82 & 0.297 & $<0.16$ & 0.0 &
$<2.96$ & $<111$ & 1.0 & 722.5/713 \\

PKS~0208-512 & RL & 1.69$\pm$0.04 & 9.49 & 31.3 &
4.6$^{+1.3}_{-1.2}$ & 32.2 & $<4.02$ & $<70.3$ & 0.0 & 201.5/268 \\

QSO~0215-504 & RQ & \ & $<0.081$ & $<2.2$ & \ & \ & \ & \ & \ &
\ \\

PHL1377 & RL & 1.72$\pm$0.06 & 2.55 & 17.0 &
3.8$^{+3.7}_{-3.3}$ & 3.0 & $<0.5$ & $<35$ & 0.0 & 224.7/253 \\

PKS~0237-233 & RL & 1.68$\pm$0.06 & 1.65 & 25.5 &
12.2$^{+6.2}_{-5.4}$ & 5.1 & $<0.26$ & $<32.2$ & 0.0 & 213.8/207 \\

PKS~0332-403 & RL & 1.71$\pm$0.12 & 1.12 & 7.75 & $<11.7$ & 1.0 &
$<2.03$ & $<325$ & 0.7 & 65.4/83 \\

NRAO\ 140 & RL & 1.67$\pm$0.02 & 10.8 & 64.5 &
10.2$^{+1.5}_{-1.8}$ & 190.4 & $<2.39$ & $<36.0$ & 0.0 & 892.7/951 \\

H~0348-120 & RL & 1.87$\pm$0.20 & 0.32 & 2.45 & $<23.0$ & 0.0 &
$<0.76$ & $<459$ & 0.0 & 47.2/50 \\

3C 109.0 & RL & 1.71$\pm$0.05 & 6.51 & 2.32 &
3.0$^{+0.7}_{-0.7}$ & 51.9 & 6.7 & 114$^{+62}_{-62}$ & 7.9 & 624.4/605
\\

PKS~0420-01 & RL & 1.75$\pm$0.07 & 2.85 & 7.96 & $<2.28$ & 0.2 &
$<3.73$ & $<224$ & 0.0 & 175.7/164 \\

PKS~0438-436 & RL & 1.63$\pm$0.13 & 1.46 & 38.7 &
22.3$^{+22.4}_{-19.9}$ & 3.4 &
$<0.99$ & $<154$ & 0.0 & 123.2/110 \\

PKS~0440-285 & RL & 1.61$\pm$0.07 & 1.27 & 16.1 & $<7.7$ & 1.4 &
$<0.78$ & $<295$ & 1.9 & 202.63/176 \\

X~0443-282 & RL & 2.04$\pm$0.13 & 7.69 & 0.22 & $<7.0$ & 0.0 &
$<8.31$ & $<477$ & 2.9 & 177.9/190 \\

E~0449-183 & RQ & 1.64$\pm$0.06 & 1.25 & 0.436 & $<0.82$ & 0.0 &
$<1.12$ & $<135$ & 0.0 & 161.9/186 \\

PKS~0528+134 & RL & 1.64$\pm$0.06 & 3.81 & 49.1 &
42.0$^{+9.0}_{-9.0}$ & 23.1 & 2.33 & 119$^{+58}_{-58}$ & 4.1 &
289.7/260 \\

PKS~0537-286 & RL & 1.47$\pm$0.08 & 2.04 & 49.1 &
14.5$^{+12}_{-10}$ & 3.0 & $<1.42$ & $<176$ & 1.4 & 152.8/163 \\

PKS~0558-504 & RL & 2.20$\pm$0.07 & 29.9 & 1.19 & $<0.2$ &
0.0 & 7.5 & 61$^{+43}_{-43}$ & 4.8 & 1254.6/1229 \\

S5~0836+715 & RL & 1.41$\pm$0.03 & 18.5 & 216 &
9.0$^{+2.5}_{-2.5}$ & 40.5 & $<5.37$ & $<67$ & 2.6 &
567.5/633 \\

3C 215.0 & RL & 1.86$\pm$0.03 & 3.46 & 1.72 &
$<0.47$ & 0.0 & 1.8 & 90$^{+57}_{-57}$ & 2.5 & 411.6/410 \\

IRAS~0910+410 & RQ & 1.91$\pm$0.06 & 2.72 & 1.78 &
4.07$^{+0.80}_{-0.74}$ & 29.3 & 8.18 & 500$^{+190}_{-118}$ & 17.5 &
452.9/379 \\

HE~1029-1401$^g$ & RQ & 1.83$\pm0.05$ & 30.9 & 0.620 &
$<0.33$ & 0.0 & 15.7 & 97$^{+40}_{-40}$ & 12.7 &
1251.7/1327 \\

Q~1101-264 & RQ & 1.84$\pm$0.24 & 0.40 & 6.26 & $<40.0$ & 0.7 &
$<1.08$ & $<487$ & 0.0 & 34.1/50 \\

HE~1104-285 & RQ & 2.10$\pm$0.09 & 1.1 & 22.8 &
18.1$^{+11.4}_{-10.1}$ & 8.2 & $<0.37$ & $<53$ & 0.0 & 205.5/199 \\

PG~1114+445 & RQ & 1.71$\pm$0.06 & 3.09 & 0.228 &
6.7$^{+0.5}_{-0.5}$ & 209.3 & 3.81 & 137$^{+76}_{-72}$ & 8.2 &
578.4/583 \\

PG~1116+215 & RQ & 2.09$\pm$0.05 & 8.12 & 0.658 &
0.43$^{+0.25}_{-0.25}$ & 5.6 & 10.4 & 274$^{+155}_{-140}$ & 9.9 &
382.6/405 \\

PG~1148+549 & RQ & 1.88$\pm$0.10 & 0.45 & 1.33 & $<3.4$ & 0.0 &
$<1.44$ & $<527$ & 0.0 & 109.7/103 \\

PG~1211+143 & RQ & 2.06$\pm$0.05 & 7.01 & 0.113 & $<0.39$ & 0.0 &
7.44$^{\it g}$ & 233$^{+72}_{-73}$ & 9.4 & 846.6/759 \\

PG~1216+069 & RQ & 1.57$\pm$0.09 & 3.59 & 1.13 & $<3.5$ & 0.0 &
2.87 & $<141$ & 0.0 & 234.6/227 \\

Mrk\ 205 & RQ & 2.13$\pm$0.03 & 15.4 & 0.184 &
0.25$^{+0.09}_{-0.09}$ & 11.8 & 12.2 & 157$^{+76}_{-79}$ & 6.5 &
986/854 \\

3C\ 273 & RL & 1.55$\pm$0.01 & 230 & 17.9 &
0.38$^{+0.03}_{-0.03}$ & 157.5 & 30.3 & 32$^{+17}_{-17}$ & 8.3 &
2034/2030 \\

PG~1247+268 & RQ & 1.83$\pm$0.09 & 0.64 & 9.21 & $<5.5$ & 0.0 &
$<0.99$ & $<304$ & 0.0 & 234.6/227 \\

MS~1248+5706 & RQ \\

3C\ 279 & RL & 1.83$\pm$0.02 & 11.6 & 10.3 &
0.52$^{+0.12}_{-0.12}$ & 19.0 & $<1.91$ & $<28$ & 0.0 & 997.4/1018 \\

B2~1308+326 & RL & 1.85$\pm$0.14 & 0.97 & 3.10 &
3.7$^{+3.4}_{-2.7}$ & 3.4 & $<1.43$ & $<251$ & 0.8 & 208.2/163 \\

IRAS~13349+2438$^g$ & RQ & 2.36$\pm$0.05 & 12.15 & 0.285 &
0.78$^{+0.23}_{-0.24}$ & 27.6 & 8.75 & 205$^{+133}_{-137}$ & 5.6 &
531.7/607 \\

RD J13434+0001 & RQ & 1.9$\pm$0.4 & 0.17 & 3.96 &
48$^{+65}_{-41}$ & 4.7 & 0.38 & 432$^{+435}_{-358}$ & 2.0 & 15.0/21 \\ 

PG~1404+226$^g$ & RQ & 1.77$\pm$0.15 & 1.5 & 0.02 & $<1.2$ & 0.0 &
3.11 & 604$^{+472}_{-445}$ & 4.6 & 160.2/175 \\

PG~1407+265 & RQ & 2.03$\pm$0.04 & 2.57 & 7.52 &
0.61$^{+0.40}_{-0.40}$ & 3.0 & $<0.95$ & $<70$ & 0.0 & 484.5/480 \\

PG~1416-129 & RQ & 1.78$\pm$0.02 & 14.1 & 0.682 & $<0.44$ & 0.0 &
12.5 & 140$^{+75}_{-75}$ & 8.1 & 650.5/636 \\

B~1422+231 & RL & 1.68$\pm$0.14 & 1.05 & 35.5 & $<151$ & 1.0 &
$<1.14$ & $<263$ & 0.0 & 87.04/75 \\

PG~1425+267 & RL & 1.72$\pm$0.07 & 3.15 & 1.40 &
3.2$^{+0.6}_{-0.6}$ & 56.8 & 2.83 & 124$^{+56}_{-76}$ & 6.1 &
460.6/399 \\

Mrk\ 478$^g$ & RQ & 1.98$\pm$0.03 & 5.75 & 0.077 &
0.66$^{+0.34}_{-0.30}$ & 5.9 & 5.36 & 191$^{+119}_{-119}$ & 5.4 &
597.7/541 \\

Q~1508+571 & RL & 1.43$\pm$0.08 & 1.03 & 40.2 & $<47.7$ & 1.0 &
$<0.51$ & $<156$ & 0.0 & 169.9/175 \\

PKS~1510-089 & RL & 1.29$\pm$0.05 & 10.1 & 4.29 & $<1.16$ & 1.8 &
6.69 & 88$^{+60}_{-60}$ & 4.9 & 547.1/540 \\ 

Q~1559+089 & RQ & 1.92$\pm$0.41 & 0.20 & 3.6 & $<31.0$ & 0.0 &
$<0.50$ & $<477$ & 0.0 & 20.96/22 \\\hline

\end{tabular}
\end{table*}
\newpage
\begin{table*}
\begin{tabular}{llccccccccc}
Table 2 continued \\
\hline
Quasar & Radio & $\Gamma$ & \multicolumn{2}{c} {Continuum$^a$} &
{\it N}$_{\rm H}^b$ & $\Delta\chi^{2}$$_{c}$ & Line Flux$^d$ & EQW &
$\Delta\chi^{2}$$_{e}$ &
$\chi^2/\nu^f$ \\
Name & Class & \ & Flux & L$_{2-10}$ \\\hline
 
PKS~1614+051 & RL & 1.43$\pm$0.14 & 0.350 & 9.41 & $<25.0$ & 0.0 &
$<0.17$ & $<132$ & 0.0 & 36.9/41 \\

4C\ 38.41 & RL & 1.60$\pm$0.19 & 1.69 & 17.0 &
11.3$^{+13.2}_{-9.3}$ & 3.5 & $<2.85$ & $<327$ & 0.0 & 38.8/51 \\

PG~1634+706 & RQ & 1.89$\pm$0.04 & 1.55 & 9.64 & $<4.7$ & 0.0 &
$<0.59$ & $<35$ & 0.0 & 294.9/276 \\

PG~1718+481 & RL & 1.94$\pm$0.04 & 2.07 & 7.93 &
$<3.2$ & 0.0 & $<0.96$ & $<88$ & 0.0 & 269.5/300 \\

QSO~1725+503 & RQ & \ & $<0.11$ & $<4.8$ & $<2.2$ \\

PDS 456$^{h}$ & RQ & 1.50$\pm$0.05 & 4.19 & 0.53 &
4.11$^{+1.69}_{-1.48}$ & 16.1 & $<1.37$ & $<32$ & 0.0 & 499.1/534 \\

Q~1745+624 & RL & 1.57$\pm$0.12 & 0.750 & 27.5 & $<79$ & 1.8 & $<0.53$ &
$<190$ & 0.0 & 74.16/78 \\

E~1821+643 & RQ & 1.90$\pm$0.01 & 23.4 & 5.98 & $<0.18$ & 0.0 &
22.8 & 167$^{+45}_{-45}$ & 30.8 & 1267/1266 \\

HS~1946+765 & RQ & 1.80$\pm$0.37 & 0.25 & 5.26 & $<240$ & 1.4 &
$<0.25$ & $<221$ & 0.0 & 37/34 \\ 

PKS~2000-330 & RL & 1.78$\pm$0.12 & 0.936 & 36.0 &
26$^{+35}_{-22}$ & 3.4 & $<1.14$ & $<239$ & 0.0 & 157.8/138 \\

3C 411.0 & RL & 1.64$\pm$0.10 & 1.15 & 0.74 & $<0.95$ &
0.0 & $<2.84$ & $<388$ & 0.3 & 80.3/79 \\

4C\ 74.26 & RL & 2.11$\pm$0.09 & 28.7 & 1.02 &
2.18$^{+0.20}_{-0.20}$ & 322.9 & 12.1 & 54$^{+50}_{-37}$ & 5.1 &
1211.0/1175 \\ 

PKS~2126-158 & RL & 1.58$\pm$0.07 & 7.97 & 253 &
14.8$^{+8.7}_{-7.2}$ & 12.5 & $<1.01$ & $<32$ & 0.0 & 232.3/250 \\

PKS~2149-306 & RL & 1.49$\pm$0.04 & 14.05 & 203 &
6.3$^{+2.0}_{-1.8}$ & 35.0 & $<1.25$ & $<22$ & 0.0 & 634.2/621 \\

PHL 5200 & RQ & 1.05$\pm$0.30 & 0.43 & 3.45 & \ & \
& \ & \ & \ &  8.6/10 \\

CTA\ 102 & RL & 1.46$\pm$0.06 & 4.04 & 13.3 &
2.8$^{+2.6}_{-2.2}$ & 4.1 & $<0.78$ & $<31.6$ & 0.0 & 194.0/206 \\

PKS~2251+113 & RL & 0.95$\pm$0.24 & 1.65 & 0.55 &
$<1.4$ & 2.0 & $<4.68$ & $<341$ & 0.0 & 139.1/106 \\

MR~2251-178 & RQ & 1.71$\pm$0.04 & 67.8 & 0.915 & $<0.06$ & 0.0 &
29.9 & 79$^{+52}_{-52}$ & 5.2 & 824.3/890 \\\hline

\end{tabular}
\end{table*}

The results from the spectral fitting are presented in table 2.
The flux is given over the
range 0.5 keV to 10 keV (in the observed \asca\ frame and extrapolated
where necessary), 
with luminosity quoted from 2-10 keV (corrected for absorption) 
in the quasar rest-frame. The value
quoted
for \nh\ is the measured value of intrinsic absorption in the quasar rest
frame, in excess of any Galactic absorption. The equivalent width (EQW) and
flux of the iron fluorescence line are given, with the EQW being in the
quasar rest frame. The line parameters are those corresponding to a
narrow line ($\sigma=0.01$~keV); the line energy is included as a free
parameter in the model fit unless otherwise stated in the table. Note
also that the photon indices $\Gamma$ 
are those of the underlying power-law continuum,
with spectral features such as low energy absorption (cold or ionised),
soft
X-ray excess, iron line emission or Compton reflection, taken into account
where these features are significant. When the fitted spectrum is
particularly complex, the 2-10 keV band index has been used. However
in most cases there is little difference between the {\it underlying}
0.6-10 keV index and the 2-10 keV index. 
All errors in the table are quoted at 68\% confidence, for the
appropriate number of interesting parameters.

For each quasar the best fitting value for reduced chi-squared
($\chi^{2}$/degrees of freedom) is given. The values of
$\Delta\chi^{2}$ quoted in table 2 are
for adding an additional spectral feature to the fit, such as an
intrinsic absorption column or Fe line. For adding an absorption column, 1
additional parameter is added to the fit and for an
iron line, 1 or 2 parameters are added, depending on whether the line energy
is fixed at 6.4 keV or not (i.e. the line normalization and line energy).
As a rough guide, $\Delta\chi^{2}=2.7$
corresponds to 90\% significance for the addition of 1 interesting
parameter,
with $\Delta\chi^{2}=4.6$ corresponding to 90\% significance for 2
additional
interesting parameters. Only measurements of \nh\ and the iron line which
are
at 90\% significance or better have been quoted as best fit values in table
3.2; otherwise upper-limits only are quoted.

F-tests have also been performed (see Bevington \& Robinson 1992), both on
the
intrinsic absorption and the iron line, to test the significance of these
features. If the value obtained for F\ $>3.0$, for the addition of 1 extra
interesting parameter, then the result has significance greater than 
$\sim$90\%. This is used as a criterion for
including line or absorption features in the spectral fitting. Actual
spectral features such as an Fe line or an absorption column are only
regarded as significant if these F-test criteria are met.

\section{The X-ray Continuum Emission from Quasars}
\subsection{General Properties}

The X-ray emission in the sample covers a wide range of photon index,
from hard (e.g. PKS 2251+113, $\Gamma=0.95$) to soft 
(e.g. IZWI, $\Gamma=2.37$).
The distribution of 2-10 photon index for the quasars in the sample is
illustrated in figure 2 (plotted against radio-loudness). However 
the X-ray emission from the majority of quasars lies in the region
from $\Gamma=1.5-2.1$. The mean photon index for all 62 quasars is 
$\Gamma=1.76\pm0.04$, with a measured sample dispersion (1$\sigma$) of
$0.27\pm0.03$.  
Compared with typical measurement errors of the order 
$\Delta\Gamma=\pm0.05$ to 0.10, this shows
that the dispersion in the X-ray continuum emission of quasars is
significant. 


Spearman-Rank correlations have therefore been performed on the photon
index ($\Gamma$), for the 62 quasars.
The strongest correlation present is a negative trend between $\Gamma$ 
and radio-loudness R$_{L}$,
which is significant at $>$99.99\% confidence (see figure 2), 
confirming the results found previously (e.g. Wilkes \& Elvis 1987).
The X-ray photon index $\Gamma$ is also
observed to decrease with X-ray luminosity \LX\ (measured in the 2-10 keV
band, QSO rest frame) and the quasar redshift z. However these correlations
are of weaker significance (at 99.7\% and 95.2\% confidence
respectively) and probably result from the strong correlation between
$\Gamma$ and R$_{L}$.
Indeed a partial S-R test performed on the data confirms that the trend
between $\Gamma$ and R$_{L}$ is the significant correlation present. 
In addition \RL\ and \LX\ are strongly correlated ($>$99.99\%), in
the sense that the most radio-loud objects tend to be more X-ray
luminous. These correlations are therefore consistent with the
increasing importance of the jet, in the more core-dominated
radio-loud quasars, due to Doppler boosting, which may account for the
flatter X-ray slopes and enhance the luminosity of the RLQs as a whole.

\begin{figure}
\begin{center}
\rotatebox{-90}{\includegraphics[width=5.8cm]{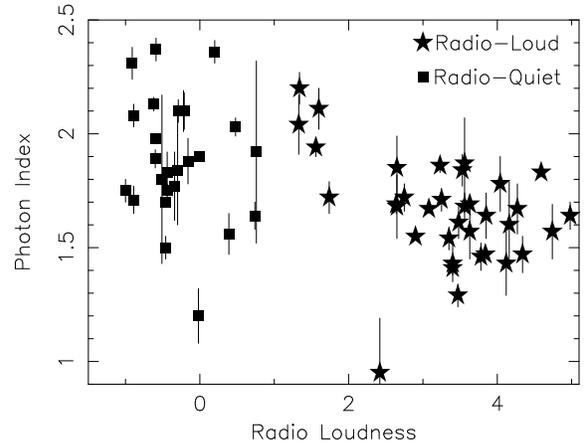}}
\end{center}
  \caption{Photon Index against Radio-Loudness, plotted for all the
  quasars in the ASCA sample.}
\end{figure}

\subsection{The Radio-Quiet sub-sample} 

We have calculated that the mean photon index for the 
{\it radio-quiet} quasars in this sample is $\Gamma=1.89\pm0.05$, 
with a significant dispersion of $\sigma=0.27\pm0.04$.
This is consistent with the mean
2-10 keV index found from an optically selected sample of PG quasars 
(George \et 2000).
Given this dispersion, correlations within just the {\it
radio-quiet} sub-sample have been investigated, excluding all the
radio-loud objects. The motivation for this is that it is possible to 
discount any effects from a
relativistic jet, and hence just learn about the properties of the quasar
central engine. No significant correlations were found between $\Gamma$ and
either \LX\ or z for the 27 radio-quiet quasars in the sample. 
Note that the spectral indices were taken over the 2-10 keV
band, in the quasar rest-frame.

Thus to 
extend the range of luminosity present, we next include the Seyfert 1s
published in the Nandra \et (1997) sample; the luminosity range covered
then extends from L$_{X}$ $\sim$ 10$^{41}$ erg/s for the least luminous
Seyfert 1 to L$_{X}$ $\sim$ 10$^{47}$ ergs/s for the most luminous quasars.
However it was found that there was still no significant
correlation between the X-ray photon index with either luminosity or
redshift. This is in contrast to the results of the Reeves \et (1997)
paper, which found a positive correlation between $\Gamma$ and \LX, but
which only considered a small sample of 9 radio-quiet quasars.
Thus this suggests that there is little evolution in the {\it underlying}
X-ray emission from radio-quiet quasars with either luminosity or redshift,
over
a wide range of luminosities and therefore presumably black hole masses. 
We have also carried out a direct comparison between our sample
of radio-quiet quasars and the sample of Seyfert 1s analysed by Nandra
\et (1997). It is found that the mean slope for the radio-quiet quasars
($\Gamma=1.89\pm0.05$) is almost identical to the mean Seyfert 1
slope (i.e. $\Gamma=1.86\pm0.05$). This also suggests that there is
little difference in the underlying X-ray emission between the Seyfert 1s
and the radio-quiet quasars.

\subsection{The Radio-Loud sub-sample}

The {\it radio-loud} quasar sub-sample has a mean index of 
$\Gamma=1.66\pm0.04$ and dispersion $\sigma=0.22\pm0.03$. 
Thus the difference in spectral slope between the
RLQs and RQQs is significant at $>$99\% confidence.
We also checked for any correlation involving $\Gamma$ within just the
radio-loud sub-sample. 
No significant correlations were found for this
sub-sample at the 99\% confidence level.

\subsection{On The Correlation between $\Gamma$ and H$\beta$ FWHM} 

What is interesting is that significant dispersion in $\Gamma$ is present
in our sample within the {\it radio-quiet quasar class alone}, that cannot
be attributed to the
properties of the radio-jet and does not seem dependent on the object
luminosity (see above). Therefore the {\it radio-quiet} 
sample was split up into
broad and narrow line objects (i.e. from the widths of the permitted
optical lines, such as H$\beta$), whereby the
narrow line objects are defined to have H$\beta$ FWHM $<2000$~km/s
(e.g. Osterbrock \& Pogge 1985). This gives a sample of 8
`narrow-lined' QSOs. It is found that the
sample of 8 narrow-line QSOs have 2-10 keV spectra that are significantly
steeper ($\Gamma=2.14\pm0.07$) than the other normal broad-lined
radio-quiet QSOs 
($\Gamma=1.81\pm0.05$). This is analogous to the difference between
narrow and broad-line Seyfert 1s found previously in the 2-10 band
(e.g. Brandt \et 1997, Vaughan \et 1999), and also in soft X-rays
(Laor \et 1997). Additionally most of the narrow line QSOs have
strong soft excesses apparent in their spectra (see later).

\begin{figure}
\begin{center}
\rotatebox{-90}{\includegraphics[width=5.8cm]{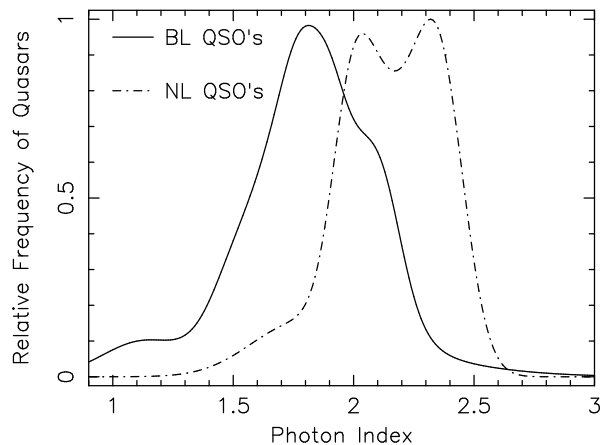}}
\end{center}
  \caption{Relative frequency distribution of 2-10 photon index for both
the broad-line QSOs and narrow-line QSOs. It is clear from the plot
that as a sample, the 8 narrow-line objects tend to have steeper
underlying spectra than the broad-line QSO, by
$\Delta\Gamma\approx0.3$. Note that only radio-quiet objects have been
used for this plot, as per text.}
\end{figure}

\begin{figure}
\begin{center}
\rotatebox{-90}{\includegraphics[width=5.8cm]{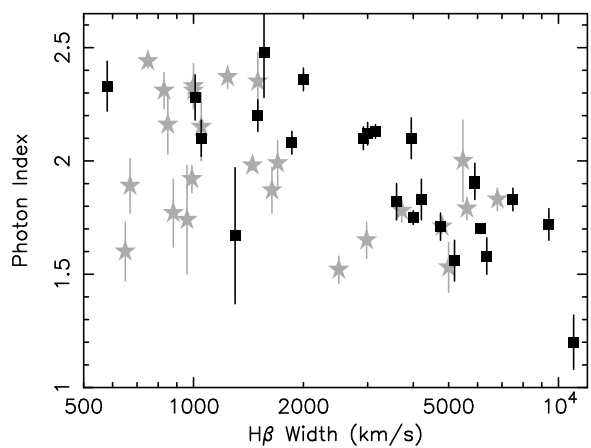}}
\end{center}
  \caption{Photon Index plotted against optical H$\beta$ FWHM width. The
high luminosity objects are shown in black and the low luminosity objects
are in grey. The trend for the photon index to decrease with increasing
H$\beta$ width is clearly seen. Interestingly the trend is not only seen in
the higher luminosity quasars, but statistically the correlation appears to
be stronger in this band.} 
\end{figure}
 
So to investigate this trend further, we performed Spearman-rank
correlations between $\Gamma$ and H$\beta$ FWHM on all the radio-quiet
quasars in this sample with known H$\beta$ widths (21 objects). The
correlation was performed using the 2-10 keV photon indicies, fitted
in the quasar rest-frame. It was
found that the photon index for quasars increases with decreasing H$\beta$
width; this yields a correlation coefficient of r=-0.61, significant at the
99.4\% confidence level for our 21 radio-quiet quasars. In order to extend
the number of objects and also to increase the range of luminosity, we next
include both the broad and narrow-lined Seyferts from the samples of Nandra
\et (1997) and Vaughan \et (1999). The total sample of objects have 
subsequently
been split into 2 bins consisting of (i) low luminosity AGN (24 objects,
L$_{2-10}=10^{42}-10^{44}$~erg/s) and (ii) high luminosity quasars
(25 objects, L$_{2-10}=10^{44}-10^{46}$~erg/s). The results of the
Spearman-Rank analysis is summarised in table 3 and the correlation is
plotted in figure 4. 

Firstly for all of the 49 radio-quiet objects, we find a strong negative
correlation between $\Gamma$ and H$\beta$ at $>99.9\%$ significance,
consistent with the results found in previous samples (e.g. Boller, Brandt \&
Fink 1996, Brandt \et 1997, Leighly 1999). Interestingly when the 2
different luminosity sub-samples are used, a significant correlation is
{\it not} found within the low luminosity bin, but a strong correlation
{\it is} found for the high luminosity objects (at $>$99.9\% significance).
There does not seem an obvious reason to explain the lack of a correlation
at low luminosites, this may just be a selection effect. Also note that the
flattest radio-quiet quasar in this sample (PHL 909, with
$\Gamma\sim1.1\pm0.1$) also has the widest H$\beta$ line profile (FWHM =
11000 km/s) of all the objects considered. However the removal of PHL
909 makes no difference to any of the correlations. {\it The important
finding here is that the apparent correlation between $\Gamma$ and H$\beta$
FWHM, found previously in low luminosity samples of Seyfert galaxies,
appears to extend to higher luminosities and hence the quasars in this
sample.} Thus in this sample the `narrow-line' quasars tend to have
steeper underlying (2-10~keV) X-ray spectra than the more common
broad-lined quasars. 

\begin{table}
\caption{
Spearman-Rank analysis on the Correlation between $\Gamma$ and H$\beta$.
{\it a} Correlation Coefficient:
{\it b} Significance of correlation;
{\it c} For all radio-quiet objects in this sample;
{\it d} Includes radio-quiet objects from this paper, Nandra \et 1997
and Vaughan \et 1999;
{\it e} Low luminosity radio-quiet AGN only (L$_{X}=10^{42}-10^{44}$~erg/s);
{\it f} High luminosity radio-quiet AGN only (L$_{X}=10^{44}-10^{46}$~erg/s);
{\it g} All radio-quiet AGN (L$_{X}=10^{42}-10^{46}$~erg/s).
}
\begin{center}
\begin{tabular}{lccc}\hline

Sample & No. of Objects & Coeff$^a$ & Prob$^b$ 
\\\hline

This Sample$^c$ & 21 & -0.646 & 99.6\% \\
Low Luminosity$^{d+e}$ & 24 & -0.332 & 87.8\% \\
High Luminosity$^{d+f}$ & 25 & -0.723 & $>99.9$\% \\
All Objects$^{d+g}$ & 49 & -0.522 & $>99.9$\% \\\hline

\end{tabular}
\end{center}
\end{table}

So a tempting hypothesis exists that can explain the dispersion
in quasar spectral slope, whereby the photon index is negatively correlated
with H$\beta$ FWHM. One explanation already postulated is that these softer
`narrow-line' objects are accreting at a higher fraction of
the Eddington-limit, which leads to greater Compton-cooling of the hard
X-ray emitting corona (Pounds, Done \& Osborne 1995, Laor \et
1997). This can naturally account for the steep hard X-ray power-law and
strong soft excesses (associated with the disk) generally observed in
this class of object. 
The other interesting finding in the last section was the lack of a
correlation between the photon index and the X-ray luminosity, for the {\it
radio-quiet} quasar sub-sample. The implication here is that the underlying
X-ray emission from {\it radio-quiet} AGN does not depend on the black hole
mass (hence the null correlation between $\Gamma$ and luminosity).
Instead the important factor may be the {\it fractional accretion rate}
($\dot{m}$), i.e. the ratio of the black hole mass accretion rate to the 
Eddington-limited rate. 
Thus the underlying X-ray spectra could depend on the
fractional accretion rate $\dot{m}$, whereby the objects with steeper X-ray
spectra are accreting at a higher fraction of the Eddington limit.

\section{The Soft X-ray Excess}

We have searched systematically for a soft excess in the spectra of
all the low redshift quasars
in this sample. Firstly the spectra were fitted in the harder 2-10
keV band, and then the spectra were extrapolated back to 0.6 keV (or 0.8
keV for the GIS) to see if there was any spectral curvature at lower
energies. An excess of counts below 2 keV, above that of the hard
power-law continuum, indicates the presence of a `soft excess'. 
In this case we have added a
blackbody component (in addition to the power-law and other spectral
components - such as an iron line) to parameterize the soft excess emission
and have subsequently re-fitted the spectrum. The results of this fitting
are shown in table 4.  The strength of the soft excess is given by the
parameter R, which represents the ratio of the blackbody to power-law
component in the 0.6-2 keV range; also given are the temperature of the
blackbody kT (units eV) and the improvement ($\Delta\chi^2$) in the
spectral fit from adding the blackbody component to the previous
best-fit model (2 additional free
parameters). Note that we have restricted this fitting procedure to objects
of redshift z$<$0.3, as at higher redshifts the soft excess will be shifted
out of the \asca\ bandpass.

\begin{figure}
\begin{center}
\rotatebox{-90}{\includegraphics[width=5.8cm]{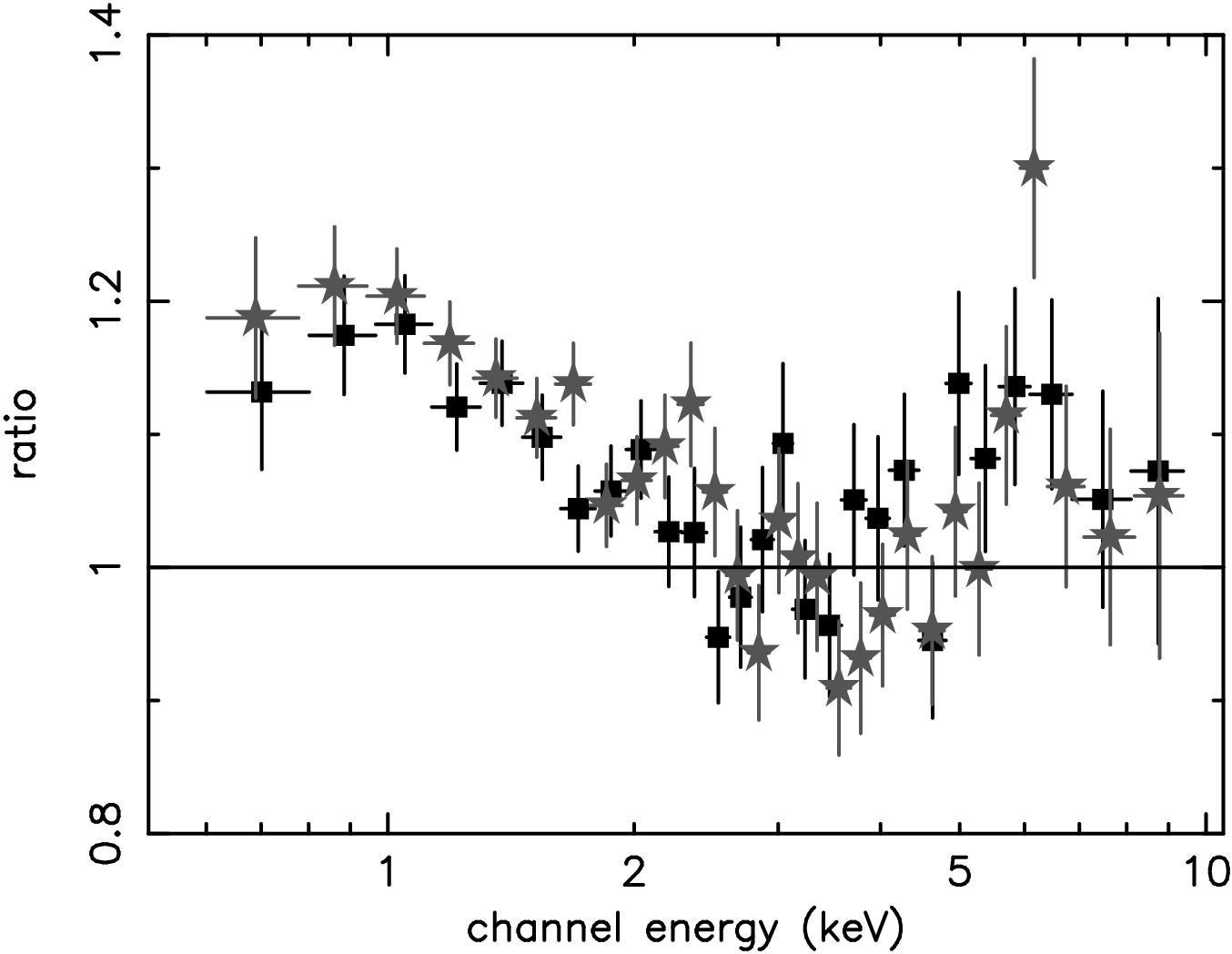}}
\rotatebox{-90}{\includegraphics[width=5.8cm]{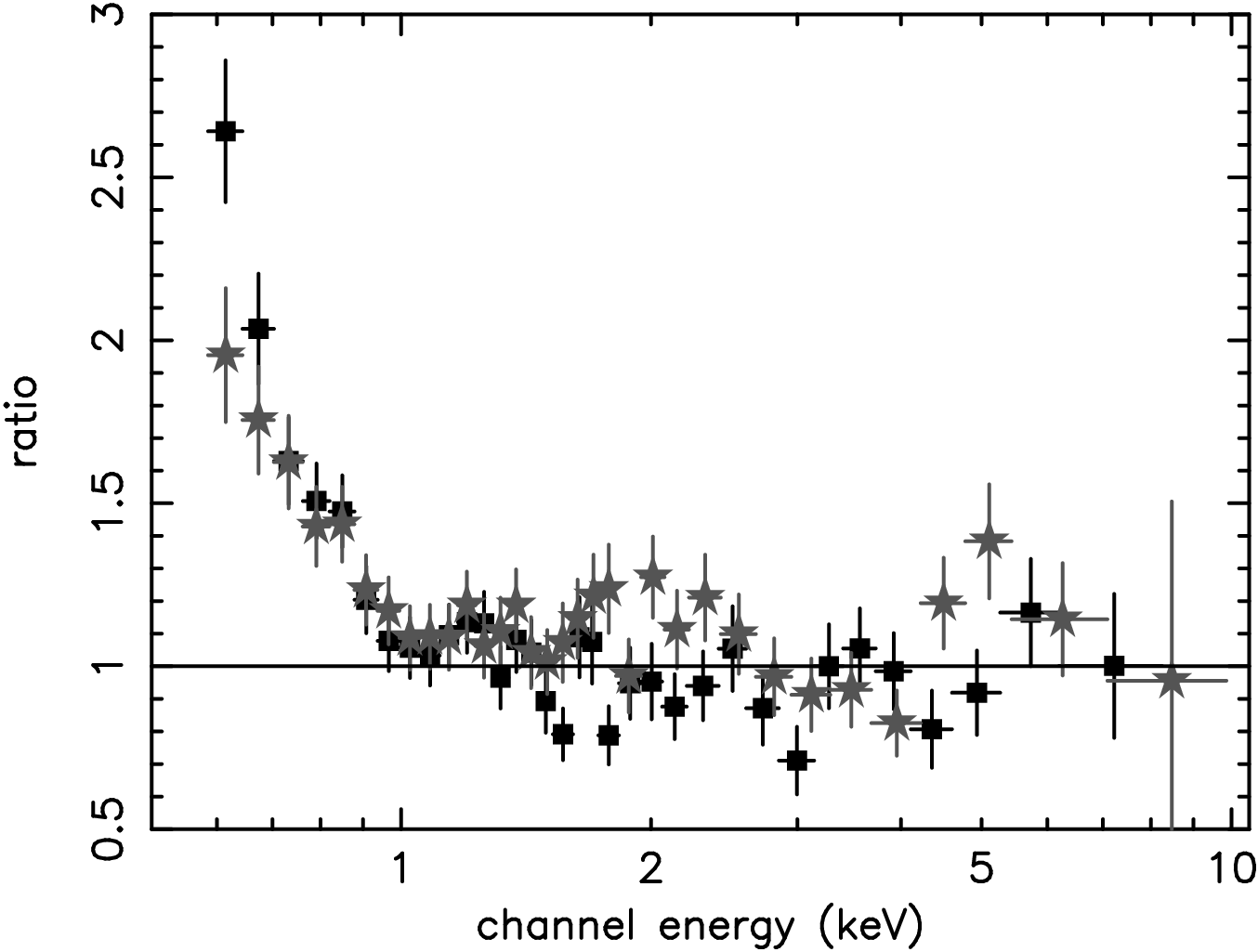}}
\end{center}
  \caption{Data/model radio residuals to SIS-0 and SIS-1 (grey) plotted for two
objects; the broad-line radio-quiet quasar HE 1029-1401 and the narrow-line
object Mrk 478. The 2-10 keV power-law has been extrapolated back to lower
energies to illustrate the spectral steepening below 2 keV. In HE
1029-1401, a gradual but significant soft X-ray excess is present below 2
keV, whilst in Mrk 478 a stronger and much steeper soft excess is present
at $<1$~keV. It is likely that whilst in Mrk 478 the soft excess could
result from thermal emission from the disk, in HE 1029-1401 an origin such
as thermal Comptonisation or ionised reflection could be more plausible
given the spectral shape.} 
\end{figure}

The table shows that some soft excess emission is significant in 9 quasars,
approximately half of the low z (z$<$0.3) objects. This soft X-ray emission
is illustrated in figure 5, which shows the data-model ratio residuals to
power-law fits in 2 objects, the broad-line QSO HE\ 1029-1401 and the
narrow-line object Markarian 478. In one object (HE 1029-1401) there is
only a gradual curvature of the spectrum below 2 keV, whereas 
the soft excess in Mrk 478 rises sharply above the power-law
continuum. Interestingly the majority of the soft excess emission occurs in
the narrow-line quasars (6 objects: TON S180, NAB\ 0205+0204,  PKS\
0558-504, PG\ 1211+143, PG\ 1404+226, Mrk\ 478), with soft excesses only
apparent in 3 broad-line quasars. This would indicate that soft excesses
are more common in the narrow-line objects, which may be expected if these
objects indeed accrete at a higher rate for a given black hole mass (e.g.
Ross, Fabian \& Mineshinge 1992).  Only one of the nine objects is
radio-loud (PKS\ 0558-504). 

The interesting question to ask here is what is the origin of the soft
X-ray excess emission? The standard explanation is that it results
from thermal 
emission that originates directly from the hot inner accretion disk 
(Malkan \& Sargent 1982) and hence is the high energy tail of the 
so-called `Big Blue Bump'. 
The temperature of this soft excess
component for the objects in this sample varies between 100 and 300 eV. In
some cases, such as in the narrow-line objects PG 1404+226 and Markarian
478, the soft excess is quite steep, with temperature $\sim100$~eV, perhaps
consistent with emission from the disk for such objects. In particular the
soft excess of PG 1404 is very strong, energetically dominating the X-ray
power-law component; this does seem to suggest that the soft excess in PG
1404+226 is a primary emission component and probably {\it does} originate
directly as thermal emission from the accretion disk.

However other cases may not be so straightforward. For instance the soft
excess in the quasar HE 1029-1401 has a blackbody temperature of
$\sim$300~eV and it is likely that temperatures of this order are perhaps
too hot to result directly from the quasar accretion disk. Note that in a
typical quasar, with a central black hole mass of  $10^8$M$_{\odot}$,
temperatures of $\ls50$ eV would be expected, implying that the observed
soft excesses in many objects are probably too hot to be the direct
emission from the putative disk. Furthermore the temperature of the disk
component varies as $T_{BB}\propto$M$_{BH}^{-0.25}$ and hence is expected
to be cooler for the more luminous objects with larger black hole masses.

Nevertheless it is plausible that some degree of Comptonisation by
electrons in a hot corona could upscatter cooler EUV photons from the disk
to soft X-ray energies and account for the observed emission. Another
possibility is that in some objects the soft X-ray emission results from
reprocessing of the hard X-ray power-law, for instance through reflection
and scattering of X-rays off the optically thick disk (e.g. George \&
Fabian 1991). Reflection will become increasingly important at soft X-ray
energies as the accretion rate $\dot{m}$ increases towards the
Eddington limit (e.g, Ross \& Fabian 1993). As the accretion rate
rises, progressively heavier
elements become fully ionised and the disk becomes more reflective at soft
X-ray energies, which can produce a steepening of the X-ray spectrum.
Furthermore ionised emission lines can be produced from abundant elements
such as O, Ne and Mg, which could also contribute towards the X-ray flux
near to 1 keV. Tentative evidence for this is present in 2 narrow-line
objects; in Ark 564 (Vaughan \et 1999b) an ionised reflection component
contributes significantly to the X-ray flux below 2 keV, whilst in the
narrow-line QSO TON S180, Turner \et (1998) report evidence for an
emission-like feature at $\sim$~1 keV. We also note that some of the
quasars in this sample (for instance TON S180, HE 1029-1401, PKS 0558-504)
may exhibit ionised iron K emission lines (see section 6), which could
provide further support of this hypothesis.

\begin{table}
\caption{
Results of fitting a blackbody soft excess to quasars.
{\it a} Ratio of blackbody to power-law component in the 0.6-2.0 keV band:
{\it b} Temperature of blackbody component (in eV);
{\it c} Improvement in the fit-statistic from adding the blackbody
component to the previous best-fit model.}

\begin{center}
\begin{tabular}{lccc}\hline

Quasar & L$_{BB}$/L$_{PL}$$^a$ & kT$^b$ & $\Delta\chi^2$$^c$ 
\\\hline

PHL\ 909 & 0.36 & 340$\pm$50 & 6.4 \\
TON\ S180 & 0.23 & 153$\pm$5 & 117.8 \\
NAB\ 0205+024 & 0.11 & 200$\pm$25 & 27.5 \\
PKS\ 0558-504 & 0.12 & 206$\pm$20 & 37.2 \\
HE\ 1029-1401 & 0.15 & 310$\pm$25 & 59.0 \\
PG\ 1211+143 & 0.09 & 125$\pm$10 & 54.0 \\
PG\ 1404+226 & 2.9 & 144$\pm$10 & 240 \\
PG\ 1416-129 & 0.19 & 307$\pm$32 & 9.5 \\
Markarian 478 & 0.27 & 107$\pm$7 & 58.9 \\\hline

\end{tabular}
\end{center}
\end{table}

\section{The Iron K$\alpha$ Fluorescence Line}
\subsection{General Properties}

Table 5 shows the list of quasars for which
Fe fluorescence line emission was found to be significant (i.e. at the
90\% level or better). The spectral fitting of the iron K lines was carried
out where
possible over the 2-10 keV \asca\ energy range.
The line parameters are first shown for a narrow line
($\sigma=0.01$~keV), with energy fixed at
6.4 keV, and then again for a narrow line but with the line energy as a
free
parameter in the fit. We have also considered broad line fine fits, where
the intrinsic velocity width of the line is a extra free parameter. However
the constraints on the line width is poor in most of the objects and so
this is not considered further.
The line parameters have all been fitted in the quasar rest
frame, i.e. corrected for redshift effects. 

\begin{table*}
\caption{
Fits to significant iron line detections.
{\it a} Units of iron line Equivalent Width (EQW) in eV.
{\it b} Units of line flux 10$^{-14}$\,erg\,cm$^{-2}$\,s$^{-1}$\,.
{\it c} Compared to a fit with no iron line
{\it d} Change due to the line energy being a free parameter in the
fit.
{\it e} Units iron line energy in keV.}
\begin{tabular}{lccccccccccc}
\hline
& \multicolumn{4}{c} {Narrow line fixed at 6.4 keV} &
\multicolumn{7}{c} {Narrow lines, line energy free parameter} \\
Object & EQW$_{a}$ & Flux$_{b}$ & $\Delta\chi^{2}$$_{c}$ &
F-prob & Line Energy$_{e}$ & EQW$_{a}$ & Flux$_{b}$ &
$\Delta\chi^{2}$$_{c}$ & F-prob & $\Delta\chi^{2}$$_{d}$ & F-prob \\\hline 

IZwI & 271$^{143}_{143}$ & 7.36 & 8.2 & 95.6\% & 
6.77$^{+0.11}_{-0.17}$ & 483$^{+212}_{-211}$ & 12.0 & 11.9 & 99.7\% &
3.7 & 94.6\% \\

PHL~909 & 158$^{+94}_{-94}$ & 5.25 & 4.5 & 96.2\% &
6.40$^{+0.18}_{-0.22}$ \\ 

TON~S180 & 96$^{+66}_{-66}$ & 3.8 & 4.9 & 97.3\% &
6.65$^{+0.10}_{-0.10}$ & 123$^{+75}_{-75}$ & 5.0 & 8.2 & 98.3\% & 4.3
& 96.1\% \\

3C~109.0 & $<55$ & $<2.7$ & 0.3 & 41.0\% & 6.78$^{+0.07}_{-0.07}$
& 114$^{+63}_{-62}$ & 6.7 & 7.9 & 98.3\% & 7.6 & 99.3\% \\

PKS~0528+134 & 119$^{+58}_{-58}$ & 2.33 & 4.6 & 95.8\% &
6.32$^{+0.16}_{-0.24}$ \\

PKS 0558-504 & $<30$ & $<3.97$ & 0.0 & 0\% &
6.66$^{+0.17}_{-0.08}$ & 61$^{+43}_{-43}$ & 7.5 & 4.8 & 91.3\% & 4.8 &
97.0\% \\
 
HE\ 1029-1401 & 60$^{+34}_{-34}$ & 10.1 & 6.8 & 99.3\% &
6.62$^{+0.16}_{-0.19}$ & 97$^{+40}_{-40}$ & 15.7 & 12.7 & 99.7\% & 5.9
& 98.7\% \\

PG\ 1114+445 & 125$^{+71}_{+72}$ & 3.5 & 7.2 & 99.3\% &
6.43$^{+0.06}_{-0.06}$ & 137$^{+76}_{-72}$ & 3.81 & 8.2 & 98.3\% & 1.0
& 68.4\% \\

PG\ 1116+215 & 145$^{+75}_{-75}$ & 6.0 & 4.3 & 96.7\% &
6.76$^{+0.08}_{-0.08}$ & 274$^{+155}_{-140}$ & 10.4 & 9.9 & 99.0\% &
5.6 & 98.5\% \\

PG\ 1211+143 & 233$^{+72}_{-73}$ & 7.4 & 9.4 & 99.8\% &
6.41$^{+0.09}_{-0.16}$ \\

Mrk\ 205 & 123$^{+70}_{-70}$ & 9.6 & 5.2 & 97.7\% &
6.37$^{+0.06}_{-0.08}$ & 157$^{+76}_{-79}$ & 12.2 & 6.5 & 96.1\% & 1.3
& 74.6\% \\

3C\ 273$^g$ & 16$^{+9}_{-9}$ & 27.0 & 3.0 & 91.6\% &
6.64$^{+0.06}_{-0.08}$ & 32$^{+17}_{-17}$ & 54.0 & 8.3 & 98.4\% & 5.3
& 97.8\% \\

IRAS~1334+243 & $<130$ & $<6.4$ & 0.4 & 50.1\% & 
6.76$^{+0.29}_{-0.21}$ & 205$^{+133}_{-137}$ & 8.7 & 5.6 & 91.3\% &
5.2 & 98.5\% \\

PG\ 1404+226 & 604$^{+472}_{-445}$ & 3.11 & 4.6 & 97.5\% &
6.2$^{+0.9}_{-0.2}$ \\

PG\ 1416-129 & 119$^{+69}_{-71}$ & 10.9 & 6.5 & 98.9\% &
6.54$^{+0.16}_{-0.18}$ & 140$^{+75}_{-75}$ & 12.5 & 8.1 & 98.3\% & 1.4
& 76.1\% \\

PG\ 1425+267 & 113$^{+55}_{-55}$ & 2.6 & 5.0 & 96.2\% & 
6.36$^{+0.06}_{-0.06}$ & 124$^{+56}_{-76}$ & 2.8 & 6.1 & 97.0\% & 1.1
& 67.1\% \\

Mrk\ 478 & 183$^{+115}_{-115}$ & 5.1 & 5.6 & 97.5\% &
6.37$^{+0.07}_{-0.07}$ & 191$^{+119}_{-119}$ & 5.4 & 5.9 & 97.0\% &
0.3 & 40.0\% \\

PKS~1510-089 & $<67$ & $<5.2$ & 0.4 & 47.0\% &
6.92$^{+0.12}_{-0.14}$ & 88$^{+60}_{-60}$ & 6.7 & 4.9 & 91.6\% & 4.5 &
96.5\% \\ 

E\ 1821+643 & 88$^{+25}_{-25}$ & 11.7 & 12.6 & $>99.9$\% &
6.61$^{+0.06}_{-0.06}$ & 167$^{+45}_{-45}$ & 22.8 & 30.8 & $>99.99$\%
& 18.2 & $>99.99$\% \\

4C\ 74.26 & 54$^{+50}_{-37}$ & 12.1 & 5.1 & 97.4\% &
6.4$^{+0.1}_{-0.6}$ \\

MR\ 2251-178 & $<36$ & $<17.3$ & 0.0 & 0.0\% & 6.57$^{+0.09}_{-0.07}$
& 79$\pm$52 & 37.3 & 5.2 & 90.9\% & 5.2 & 98.2\% \\\hline

\end{tabular}
\end{table*}

The F-test and F-distribution (Bevington \& Robinson 1992) were used to
test
the significance level of any line features, the results of which are also 
shown in the table. The change in $\Delta\chi^{2}$ for a fit
with narrow line fixed at
6.4 keV, compared with a fit with no Fe line, is quoted along with the
associated probability from performing an F-test with 1 additional
parameter. Then the fit to a narrow line with the line energy free is
compared to one with no lines, the values for both $\Delta\chi^{2}$
and the F-test
probability (for 2 additional parameters; the line normalisation and the
line energy) are both quoted. Also quoted is the additional probability
(and
$\Delta\chi^{2}$) for freeing the line energy in the fit; i.e. the fit
to a line with
free energy is compared with a line fixed at 6.4~keV. This then can be used
as a measure of whether the line energy is significantly different to the
neutral value of 6.4 keV. 
Line emission is deemed significant if the F-test probability is $>$ 90\%;
in
some cases lines are not significant at 6.4 keV, but are when the line
energy is
free to vary, indicating that the emitted line energy is different to
6.4~keV.

Iron line emission has been detected in 21 out of the final
62 objects in the sample. There is a large spread in the best-fit value of
line equivalent width (EQW), ranging
from 32eV for 3C 273 up to $>$400eV for the radio-quiet objects IZwI and PG
1404+226, although the main peak of
quasars with detected emission between 50 and 200eV is apparently
consistent with Compton reflection origins (e.g. George \& Fabian 1991).
In the case of the radio-quiet quasars there is evidence of Fe K$\alpha$
line features, with 14 out of 27
objects showing evidence for line emission above the 90\%
confidence level. At the 99\% (or better) confidence level, Fe K emission
lines are detected in IZwI, TON S180, HE 1029-1401, PG 1114+445, PG
1116+215, PG 1211+143, Mrk 205 and E 1821+643. 
Interestingly lines are not detected in the
most luminous (L$_{X}>10^{46}$ ergs/s) of the
radio-quiet quasars at higher redshifts; in particular PG 1634+706
($<$35eV), HE 1104-285 ($<$53eV), PG 1407+265 (EQW$<$70eV) and PG 1718+481
($<$88eV). 

In contrast to the radio-quiet quasars, significant line emission has been
detected in only 7 of the 35 radio-loud quasars. Generally, the line EQW
found here is smaller than that for the radio-quiet objects, with
EQW$<$100eV in many cases. Where no line emission has been detected, tight
upper limits have been placed on the objects in several cases.
For example, the blazars 3C 279 and CTA 102
have line EQWs $<$28eV and $<$32eV respectively, whilst the distant
radio-loud quasars PKS 2126-158 and PKS 2149-306 also have tight limits of
$<$32eV and
$<$22eV respectively. Most of these quasars with low upper-limits on the
line EQW are core-dominated quasars. 

\begin{figure}
\begin{center}
\rotatebox{-90}{\includegraphics[width=5.8cm]{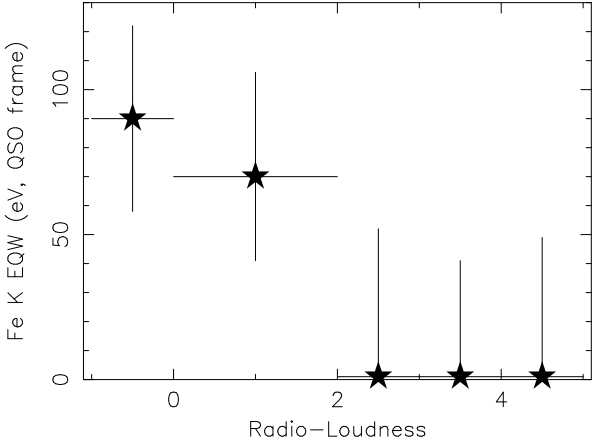}}
\end{center}
  \caption{Mean iron K$\alpha$ line equivalent width plotted against
  quasar radio-loudness. The trend is for the line strength to
  diminish with increasing R$_{L}$, as Doppler boosting of the
  continuum by the jet becomes increasingly important.}
\end{figure}

As a whole we find that the mean line equivalent width (detections
only, narrow line fits) in the RQQs is $163\pm17$ eV, 
whereas for RLQs the mean is $85\pm15$ eV. This suggests that the amount of
iron line
emission is weaker in the radio-loud objects. To investigate this further
we
performed a Spearman-rank correlation (using survival statistics - see
Isobe, Feigelson \& Neilson 1986), which showed that the line EQW
(for narrow lines) decreases with quasar radio-loudness at 99.99\%
confidence. This trend is also illustrated in figure 6, the
mean iron line EQW (including upper limits) has been 
plotted against radio-loudness, with the quasars split up into different
bins for R$_{L}$. As a whole Fe lines are
detected in the radio-quiet objects, but not in the radio-loud
quasars. The simple interpretation of this trend is that the Fe emission
diminishes as the jet angle approaches the line
of sight. Thus the increased Doppler boosting
of the X-ray continuum in relation to the line, weakens
the relative strength of the disk reflection component in
core-dominated RLQs. 

We have also investigated any trends in the line emission for the 
{\it radio-quiet} objects, excluding the radio-loud quasars. This
negates the effect that the powerful relativistic jet has on any of the
correlations. In order to extend the range in luminosity, we include
the 18 Seyfert 1s from the Nandra \et (1997) sample, as well as our own
27 RQQs. A Spearman-rank
analysis shows that the line EQW is negatively correlated with X-ray
luminosity at 99.98\% confidence, even when only the radio-quiet AGN
are considered. This correlation confirms the `X-ray Baldwin effect'
that was found previously from Ginga (Iwasawa \& Taniguchi 1993), and also
\asca\ data
(Nandra \et 1997b).
Thus the Fe line emission appears to be absent in the most
luminous of AGN, regardless of whether they are radio-loud or quiet. 

\subsection{Iron Line Energy}

Overall it is found that in 11 out of 21 of the quasars that show 
Fe K$\alpha$ line emission, the
line emission is at rest energies $>$6.4 keV at 90\% confidence or
better. Thus in at least half of the quasars in the sample, 
the line originates from matter that is partly
ionised. For the RQQs, 7 objects have partially ionised Fe K$\alpha$ line
emission; these are IZwI, TON S180, HE 1029-1401, PG 1116+215, IRAS
13349+2438, E 1821+643 and MR 2251-178. For the radio-loud
quasars, 3C 109.0, PKS 0558-504, 3C 273 and PKS 1510-089 appear to have Fe
lines that originate from partially ionised material. 

The overall range of line energy, from 6.4 keV to 6.9 keV,
represents emission from a variety of ionisation states from
neutral iron (Fe \textsc{i} to Fe \textsc{xvi} at $\sim$6.4 keV) 
up to helium or even hydrogen-like iron (Fe \textsc{xxv} at 6.68 keV, 
Fe \textsc{xxvi} at 6.96 keV). 
This is in contrast to the line properties of lower luminosity
AGN such as Seyfert1 galaxies, where the Fe K$\alpha$ emission
normally originates
from neutral iron, with the line energies closely distributed near to
6.3-6.4 keV (Nandra \et 1997). For instance the mean line energy of the
Nandra \et (1997) sample of Seyfert 1 galaxies is $6.37\pm0.02$~keV,
whereas for our quasar sample the mean line energy is
6.62$^{+0.05}_{-0.07}$~keV. A Spearman-rank correlation also shows that the
line energy increases with object luminosity at $>$99.9\% significance
(also see figure 7), again providing strong confirmation of the Nandra \et
(1997b) result whereby the iron lines in high luminosity sources tend be
weaker and shifted bluewards of 6.4 keV.

\begin{figure}
\begin{center}
\rotatebox{-90}{\includegraphics[width=5.8cm]{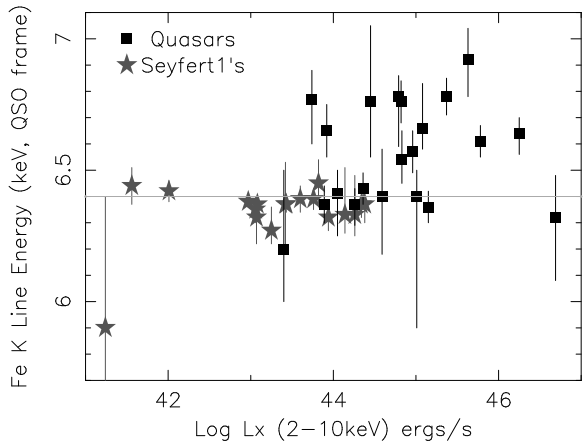}}
\end{center}
  \caption{Iron K line energy against X-ray luminosity, plotted for both
  quasars (squares) and Seyfert 1s (stars, grey). The iron
  line energy (and hence ionisation state) increases with luminosity,
  indeed many of the quasars have energies significantly $>$6.4 keV.}
\end{figure}

\subsection{The Compton Reflection Hump} 

Having ascertained that there is substantial iron line emission from
several of the quasars in this sample, we next determine the
properties of the Compton reflection hump that would be expected to
accompany the line, on the assumption that 
the line results from reflection off optically
thick matter. Constraints have therefore been placed on
the amount of Compton reflection that occurs in the spectra of quasars
in this sample. This is particularly important because if the iron
lines that are described above result from the accretion disk, then we
would expect to see evidence for a reflection `hump' in the spectra of
quasars. To test this, 
a Compton reflection model has been used (the \textsc{pexrav} code in
XSPEC, Magdziarz \& Zdziarski 1995) with solar
abundances, a disk inclination angle of 30 degrees has been assumed
(unless otherwise stated) and an exponential cut-off at $>100$~keV. 
The strength of the reflection component (R) is measured in
terms of the solid angle $\Omega$ subtended by the primary X-ray source to
the disk; then R is then given by R $=\Omega/2\pi$. 
Although R is left as a free 
parameter in the spectral fitting, normally for a geometrically thin
disk a value of R near to 1 would be
expected. The model also used assumes that the `reflecting' material
(in this case presumably the accretion disk) is of a neutral or low
ionisation state.

\begin{table*}
\caption{
Results of fitting Compton reflection to quasars. 
{\it a} X-ray luminosity measured in 2-10 band (QSO frame), units
erg/s;
{\it b} Strength of reflection component (R~$=\Omega/2\pi$). Errors
are at 68\% for 2 interesting parameters;
{\it c} 90\% Upper limits on R;
{\it d} Change in $\chi^{2}$ statistic when R is fixed at 1.
A positive number corresponds to the fit statistic becoming worse.}

\begin{tabular}{lcccccc}\hline

Quasar & Type & z & Log L$_{X}$$^{a}$ & R(=$\Omega/2\pi$)$^{b}$ & 90\%
UL$^{c}$ & $\Delta\chi^{2}_{R=1}$$^{d}$ \\\hline

S5~0014+813 & RL & 3.41 & 47.0 & $<0.02$ & $<0.03$ & +133.2 \\
NRAO~140 & RL & 1.258 & 46.76 & $<0.05$ & $<0.08$ & +147.5 \\
S5~0836+716 & RL & 2.17 & 47.3 & $<0.07$ & $<0.09$ & +162.4 \\ 
HE 1104-1805 & RQ & 2.319 & 46.4 & $<0.09$ & $<0.15$ & +21.3 \\
3C~273 & RL & 0.158 & 46.25 & $<0.25$ & $<0.33$ & +57.2 \\
PG 1247+268 & RQ & 2.041 & 46.0 & $<0.25$ & -- & +3.0 \\
Q 1508+571 & RL & 4.3 & 46.6 & $<0.1$ & $<0.2$ & +16.3 \\
PG 1634+706 & RQ & 1.334 & 46.0 & $<0.275$ & $<0.38$ & +17.5 \\
PG 1718+481 & RQ & 1.084 & 45.9 & $<0.15$ & $<0.25$ & +26.0 \\
HS 1946+765 & RQ & 3.02 & 45.7 & $<0.2$ & -- & +5.2 \\
4C 74.26 & RL & 0.104 & 45.0 & 3.0$\pm$1.4 & -- & -39.5 \\
PKS 2126-158 & RL & 3.26 & 47.4 & $<0.025$ & $<0.05$ & +68.0 \\
PKS 2149-306 & RL & 2.345 & 47.3 & $<0.06$ & $<0.08$ & +175.5 \\\hline 

\end{tabular}

\end{table*}

The Compton reflection component was systematically fitted to the
brightest quasars in the sample (both at low and high z), i.e. where
the signal-to-noise ratio allows a sufficient constraint of these
additional model parameters. 
The results are now presented here and are shown in
table 6 for those quasars in which the reflection component has been
constrained; the table either shows the best fitting value of the
relative strength, R, of the reflection component or the statistical
upper limit to R.  

It can be seen that in the low redshift quasars it has generally 
not been possible to constrain the
amount of reflection present in the X-ray spectra. The only low
redshift quasars in which the amount of reflection has been
constrained are in the radio-loud quasars 3C 273 and 4C 74.26. The 68\%
upper limit on the strength of the reflection component in 3C 273 is
R $<0.25$ (or R $<0.33$ at 90\% confidence). This result is perhaps not
surprising as 3C 273 is a core dominated radio-loud quasar, the
reflection component should be considerably diluted by the beaming
effects of a relativistic jet. 
An unusually strong reflection component is found in the lobe-dominated
radio-quasar 4C 74.26 (also see Brinkmann \et 1998); 
however the presence of a strong soft excess
(with temperature KT$\sim300$~eV for a blackbody) 
could confuse the detection of this reflection component, given the
limited \asca\ bandpass. Constraints are not possible on the amount of
reflection in the other low z objects.

However, unlike for low z quasars, in the high redshift quasars there is a
greater band-pass in hard X-rays, as this ($>10$~keV) part of the
spectrum is redshifted into the \asca\ rest-frame. 
Therefore in the high redshift quasars, which have sufficient signal
to noise, it should be possible to constrain the higher energy Compton
reflection component. Firstly for the high redshift radio-loud
quasars, it can be seen from the table that the reflection component
is very weak or consistent with no reflection,
with the relative strength of the component constrained to R
$<<1$. In fact the presence of a reflection component
with R=1 (i.e. as would be expected from scattering of X-rays off an
accretion disk) is excluded in all the distant radio-loud quasars
considered here, at $>$99.99\% confidence. Examples are S5 0014+813, NRAO
140, S5 0836+715, Q 1508+571, PKS 2126-158 and PKS 2149-306. However
the lack of a reflection component in these bright radio-loud quasars
is perhaps not surprising as many of these objects are believed to be 
jet-dominated
AGN, where the Doppler boosted jet component can dominate over the
reflection component from near to the central engine; this is also
consistent with the weak iron lines that are observed
in this class of object.

It has also been possible to constrain the amount of reflection in some of
the high redshift (z~$>1$) {\it radio-quiet} quasars. In the luminous quasars
PG 1634+706, PG 1718+481 and HE 1104-1805, the reflection component is
constrained to R $<<1$. The upper limits (90\% confidence) on these 3
quasars are R $<0.38$, R $<0.25$ and R $<0.15$ respectively; in addition a
reflection component in the high z QSO HS 1946+765 is also constrained
to R $<0.2$. The presence of a reflection component with
R=1 (as expected from an accretion disk), is excluded at $>$99.9\%
confidence in PG 1634+706, PG 1718+418 and HE 1104-1805. All 3 of
these quasars are particularly luminous (with L$_{2-10}\sim10^{46}$ erg/s
and M$_{V}$ = -28 to -29, using q$_{0}$=0.5); 
both PG 1634 and HE 1104 are radio-quiet
(R$_{L}=-0.587$ and R$_{L}=-0.94$ respectively), and PG 1718 is
radio-intermediate (R$_{L}=1.58$). Compton reflection has not been
constrained in the other distant radio-quiet quasars (lack of
signal-to-noise). However the evidence suggests
that the amount of {\it neutral} reflection in highly 
luminous high redshift quasars is much weaker than what is observed in
lower luminosity Seyfert~1 galaxies (e.g. Pounds \et 1990 Nandra \&
Pounds 1994). As this trend is observed in the
radio-quiet as well as the radio-loud quasars, the lack of reflection
features in quasars cannot
just be interpreted in terms of increased Doppler boosting from a
relativistic jet. This, along with the above trends in the iron line
emission, will be discussed in the next section.

\subsection{The Nature of Iron lines and Disk Reflection in Quasars}

The analysis performed in the above section show that properties of
iron line emission and the associated reflection hump in quasars seem
less straightforward than in the Seyfert 1s. As was explained
earlier, the common picture in the Seyfert 1s is for the hard X-rays
to scatter off the inner accretion disk, producing a broad iron K$\alpha$
line, iron K edge and a Compton reflection hump. The fact that the line is
highly broadened and fits the relativistic motions expected in the
inner disk around a black hole (e.g. Tanaka \et 1995), strongly
supports the interpretation that these `reflection' features originate
from the accretion disk. In the Seyfert 1s, this
reprocessing is almost completely from material that is
neutral or of a low ionisation state (i.e. Fe \textsc{i} to Fe
\textsc{xvi}).

However this is apparently not the case in quasars. It was found
that of the 21 quasars in this sample with significant iron K line
detections, a large proportion of these (at least 11) have iron line
energies ($>6.6$~keV) that are consistent with originating from 
highly ionised material (e.g. Fe \textsc{xxiv} to Fe
\textsc{xxvi}). In the high luminosity
end of the quasar distribution (both radio-loud and radio-quiet), the
actual strength of the iron K line component decreases considerably
(i.e. from EQW $=100-150$~eV to EQW $<50$~eV), providing strong
confirmation of the results in
Nandra \et (1997b). Additionally we also find that the
strength of the {\it neutral} reflection hump is diminished considerably
in the high luminosity quasars.

So how can one explain the observations in the quasars, which seem to
contrast with the situation in the Seyfert 1 galaxies? Firstly the energy
of the line emission in these quasars implies that the ionisation
state of the reprocessing material (the surface layers of the
putative accretion disk) is clearly much higher than in the Seyfert
1s. As has been pointed out from photoionsation modeling 
(e.g. see Ross \& Fabian 1993; Matt, Fabian \& Ross 1993; 
Ross, Fabian \& Young 1999), the surface
layers of such disks can become substantially photoionised as the
{\it fractional accretion rate} ($\dot{m}$) of the central engine
increases. 
Perhaps at the Eddington limit very little if any iron K emission is seen,
as the disk becomes fully ionised down to several Thomson depths.

This picture is consistent with
the most luminous quasars (either radio-loud or radio-quiet), 
with very weak iron lines, accreting near the Eddington limit.
Additionally, in the radio-loud objects, the relativistic jet can further
weaken the line and reflection component.
Furthermore this hypothesis can also explain the apparent lack of a
reflection hump from {\it neutral} material in the high z quasars,
found in section 6.3. Indeed it has been postulated (Ross, Fabian \&
Young 1999), that the disk becomes more reflective in high luminosity
quasars (as the photoelectric opacity below Fe K decreases 
at higher ionisation) which can reproduce the apparently featureless
spectra of some high z quasars. The lack of contrast
between the continuum and the reprocessed X-rays can then explain the apparent
absence of a reflection `hump'.
Further support for the presence of a highly ionised
accretion disk in one object comes from an observation (
with \asca\ and \xte) of
the luminous quasar PDS~456. Although the line
emission is fairly weak, a deep, highly ionised edge
(at 8.7 keV) is present in the spectrum of this quasar. This iron K edge
could originate from a {\it high ionisation} reflector. 
The detailed spectral fitting of this quasar, from {\it
simultaneous} \asca\ and \xte\ data, is discussed in a separate paper
(Reeves \et 2000). 

\section{X-ray Absorption in Quasars}

\subsection{General Properties}

The intrinsic neutral column densities for the 62 quasars analyzed are
shown in
table 2, fitted in the rest frame of the quasar. An additional Galactic
absorption component was also fitted in the observers rest frame, as
described previously. Unless an
intrinsic absorption column is detected at 90\% confidence (in addition to
the Galactic absorption), an upper limit only has been quoted. Intrinsic
absorption (at $>$90\% significance) 
is present in 20 radio-loud quasars (RLQs)
and 10 radio-quiet quasars (RQQs), i.e. 30 out of the 62 quasars.

The strongest absorption is seen in the most distant radio-loud quasars.
Out
of the 20 RLQs with significant intrinsic absorption, 9 of the 13 RLQs at
z $>2$ show evidence for absorption with \nh\ in the order of 
10$^{22}-10^{23} {\rm cm}^{-2}$.
In particular, strong intrinsic absorption is seen in the distant
radio-loud
quasars S5 0014+813, PKS 0528+134, S5 0836+715, PKS 2126-158 and PKS
2149-306 all at $>$99.9\% significance and columns of about $10^{22}
{\rm cm}^{-2}$ or
greater. The largest absorption column is seen in S5 0014+813, at red-shift
z = 3.4, with a column density of ($6.5\pm2.6$)$\times10^{22} {\rm
cm}^{-2}$ in the rest frame
of the quasar (although note that the Galactic absorption towards 
S5~0014+813 is rather high). 

Considering radio-loud quasars at intermediate redshifts ($1<{\rm
z}<2$), 6 out of
nine quasars also have evidence for absorption, but with lower columns than
at high z, typically $10^{21}-10^{22} {\rm cm}^{-2}$. 
However low z radio-loud quasars at redshifts
z$<$1 generally only show weak absorption (\nh $<10^{21} {\rm
cm}^{-2}$). Significant
absorption is only seen in 5 out of 13 low z radio-loud quasars, these are
3C 109.0, 3C 273, 3C 279, PG 1425+267 and 4C 74.26. 3C 109.0 and 4C 74.26
are more lobe dominated than some of the other RLQs, which may imply some
dependence on orientation; for these two objects \nh~$\approx10^{21} {\rm
cm}^{-2}$. The columns towards 3C 273 and 3C 279 are both low with \nh
$<10^{21}{\rm cm}^{-2}$.

In comparison, significant absorption is also seen in only 10 radio-quiet
quasars out of 27. At high redshifts (z $>2$) there are only 2 radio-quiet
quasars with statistically significant intrinsic absorption; HE 1104-285
(at z=2.319) has a column of $1.81\times10^{22}{\rm cm}^{-2}$ at $>$99\%
significance, whilst RDJ 13434+0001 (at z=2.350) has an intrinsic
column of $4.8\times10^{22} {\rm cm}^{-2}$ although
this is less well constrained. HE 1104-285 is
unusual in that it is associated with a gravitational lens system
(Wisotski \et 1993), and RDJ13434+0001 may be a rare example of a type II
QSO (Almaini et al. 1995).
In the case of the low z RQQs, virtually all the
objects are consistent with columns of \nh $<10^{21} {\rm cm}^{-2}$, 
i.e. similar to Galactic sized columns.

\subsection{The Warm Absorber}

However there are some low z quasars which show evidence for absorption
from warm (or partially ionised) matter. An example of this is the
radio-quiet
quasar PG 1114+445, which has a warm absorber with a column of typically
\nh$\sim10^{21}$ to $10^{22} {\rm cm}^{-2}$, 
with the main contribution from the ionised O\textsc{vii} and
O\textsc{viii} edges. Similar results for PG 1114 were reported by George
\et (1997). Another low z warm absorber quasar similar to PG 1114+445
is MR 2251-178 (Reeves \et 1997, Pan, Stewart \& Pounds 1990). We also
detect a significant ionised absorption component in the quasar IRAS
13349+243, this has been reported previously by Brandt \et (1997b) who
suggest that a dusty warm absorber is likely.

There are only 2 other quasars, not reported in the literature to date,
with possible evidence for a warm absorber in this sample. One of these is
the radio-loud quasar PG 1425+267 which has a warm absorber of column a few
$\times10^{21} {\rm cm}^{-2}$, although the detection is marginal statistically
(95\% confidence) and the ionisation ($\xi\sim10~{\rm erg~cm~s}^{-1}$)
is rather poorly constrained. However perhaps the most
interesting case is the luminous (M$_{V}\sim-27$) nearby radio-quiet quasar
PDS 456. The complex residuals to a power-law fit for this quasar are shown
in figure 8. An iron K edge is present at  8.7$\pm$0.2~keV, and the optical
depth of the edge is very large ($\tau=1.6\pm0.5$). The edge is detected at
$>$99.99\% confidence (or $\Delta\chi^{2}=56$ for 2 interesting
parameters). In addition we can also fit a low energy warm absorber
component to PDS 456, with a high column of material ($5\times10^{22}{\rm
cm}^{-2}$) and ionisation parameter ($\sim500~{\rm erg~cm~s}^{-1}$). The
spectrum of PDS 456 is discussed in detail in a separate
paper (Reeves \et 2000). The main interpretation of the fitting is that
there is a warm absorber component at lower energies, but that a high
ionisation disk reflector could be responsible for the deep iron K edge.
There is no other evidence for the presence of such a strong iron K edge in
any of the other quasars, although the constraints in many cases are quite
poor.

\begin{figure}
\begin{center}
\rotatebox{-90}{\includegraphics[width=5.8cm]{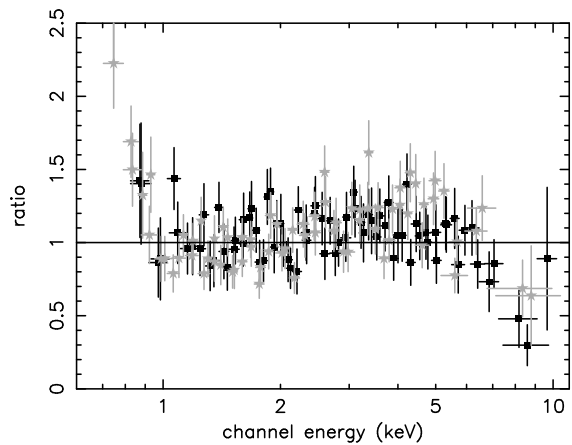}}
\end{center}
\caption{Data-model residuals to a power-law fit for the radio-quiet quasar
PDS 456. Complex features are seen, notably a deep iron K edge at 8.7 keV,
a warm absorber at low energies and some excess in soft flux. 
The edge is highly ionised, ($\xi=6000~{\rm erg~cm~s}^{-1}$) 
and could result from either a high ionisation absorber or
reflector (see Reeves \et 2000.)}
\end{figure}

On the whole in this sample there are only a few examples of warm
absorbers in quasars. This is also supported by \rosat\ PSPC
observations of PG quasars (Laor \et 1997), in which only about 5\%
have X-ray spectral features due to a warm absorber. George \et (2000)
also found relatively few warm absorbers in an \asca\ sample of PG quasars.
In contrast the occurrence of warm absorber
features is common in the Seyfert 1s (e.g. Reynolds 1997, George \et
1998). This may be due to the absorbing material being at a different
level of ionisation in the
quasars. For instance if such matter was very highly ionised, the material
would be
essentially transparent at O\textsc{vii} and O\textsc{viii}
energies. Another possibility is that there is a lack of such
material in quasars. However a more straightforward explanation may simply
be due to quasar redshift, where the O\textsc{vii} and
O\textsc{viii} features are red-shifted out of the \asca\ bandpass.

\subsection{On the correlation between X-ray absorption and redshift}
The observations that have been discussed in section 7.1 suggest that
the amount of neutral 
X-ray absorption seems to be higher in the high z objects than in
the low z quasars. We have therefore performed
Spearman-rank correlations in order to investigate any apparent
correlation between {\it neutral} \nh\ and z. 
As in previous sections, the use of survival
statistics has been employed in order to allow upper-limits to be
used in the correlations. (Note that all quasars with flux levels
$<1\times10^{-12}$~erg cm$^{-2}$ s$^{-1}$ were excluded from these
correlations, as useful constraints cannot be placed on the amount of
absorption present.) A simple correlation between \nh\ (fitted
in the quasar frame - corrected from Galactic absorption) and z is
significant at $>$99.99\% confidence; thus the amount of absorption 
{\it intrinsic} to the quasar is seen to increase with redshift. 
This apparent trend has been reported previously, with smaller samples of
objects (see Elvis \et 1994, Cappi \et 1997, Reeves \et 1997, Fiore
\et 1998). 
Typically columns in high
redshift (z$>$2) quasars are of the order $\sim10^{22}$cm$^{-2}$ 
(QSO frame), 
whereas at low redshift {\it N}$_{H}$ is only of the order few
$\times$ 10$^{20}$cm$^{-2}$. The positive
correlation between \nh\ and z is illustrated in figure 9. 

\begin{figure}
\begin{center}
\rotatebox{-90}{\includegraphics[width=5.8cm]{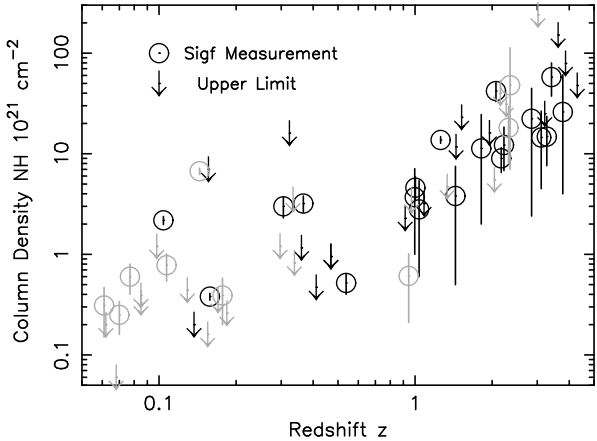}}
\end{center}
\caption{Quasar intrinsic column density (or {\it N}$_{H}$) plotted
  against redshift. The {\it N}$_{H}$ shown is corrected for Galactic
  absorption and is plotted in the QSO rest frame. Upper-limits are
  indicated by the arrows. Radio-loud objects are in black,
  radio-quiet objects are shown in grey.}
\end{figure}

It as has been reported (e.g. Cappi \et 1997, Mukai 1998) 
the average neutral
column in the SIS0 and SIS1 detectors tends to be overestimated by
$2-3\times10^{20}{\rm cm}^{-2}$, due to uncertainties in the low
energy calibration of these instruments. However it is also worth
noting that a simultaneous Beppo-Sax and \asca\ observation of 3C 273
showed that the SAX LECS and MECS instruments were in agreement with
the ASCA SIS (Orr \et 1998). 
Nevertheless, in order to take into account any possible
calibration effect, an additional column of 
\nh$=3\times10^{20}{\rm cm}^{-2}$ 
has been added to the spectral fit, in the local z=0 \asca\
rest-frame, and the intrinsic columns have subsequently been refitted. Note
that all SIS data below 0.6 keV have also been ignored, where the
calibration problems tend to be worse.  
Despite of this, the correlation between \nh\ and z was
still found to be significant at $>$99.9\% confidence.
Excluding all those quasars that lie at low Galactic
latitudes ($\beta\la20\degg$), also has little effect on
the significance level of the correlations.

By fitting the absorption in the quasar rest-frame it is possible that
the amount of \nh\ could be overestimated, as the absorbing material could
lie anywhere along the line-of-sight to the quasars. Indeed if there was
just some local effect that could produce some spurious \nh\
measurement (such as SIS calibration or uncertainty in the
Galactic column) then this would be magnified towards the quasars at
higher z. Therefore the above correlations have also been performed by
fitting the absorption column in the local z=0 frame (after correcting
for the Galactic column). The correlation between \nh\ and z is still
found to be significant at $\sim99.9$\% confidence even in this case. Again
the calibration uncertainties of \asca\ have been taken into account by
adding an additional column of $3\times10^{20}{\rm cm}^{-2}$ and by
removing all quasars at low $\beta$. However even after these checks
were performed, the correlation remained significant. 
Thus the statistical evidence, even after accounting for
calibration and uncertainties in the Galactic absorption, seems to
suggest that the correlation between \nh\ and z is real. 

Therefore the most striking observation here is the discovery of moderately
large
absorption columns in several of the high z quasars. As these high z
quasars are predominantly type I radio-loud AGN, 
obscuration in terms of the molecular torus
(as observed in Seyfert 2s) seems unlikely. The physical origins of this
absorption can then either be local to the rest frame of the quasars at
high z, or it may be associated with matter at intermediate redshifts
(i.e.
from intervening line of sight material) not physically connected with the
quasars. It is possible that the if the absorbing material is intrinsic to
these quasars, then it could be similar in origin to the high ionisation
absorbers observed in more nearby AGN. The most
substantial source of soft X-ray absorption from intervening matter would
probably be from damped Lyman-$\alpha$ absorption systems, however 
the number density of these dense systems is reported to be
quite low (O'Flaherty \& Jakobsen 1997).
A more detailed account of the possible origins of this X-ray absorption
will not be discussed further in this paper, but can be found in
the literature (e.g. Elvis {\it et al.} 1998, Cappi {\it et al.} 1997,
Reeves \et 1997, Elvis \et 1994). However
observations using the superior low energy throughput
provided by XMM and Chandra are needed to confirm this trend, and to
provide further clues as to the possible causes.

\section{Summary \& Conclusions}

We now summarize the findings of this paper. In particular, comparison is
drawn to our earlier paper (Reeves \et 1997 or R97), which contained a
smaller sample of quasars (24 objects compared with the current sample size
of 62).

Firstly we confirm the following main results from the R97 paper:-

\begin{itemize}

\item{A decrease in the X-ray photon index, with radio-loudness, 
for all quasars.}

\item{A strong confirmation of the correlation between the neutral X-ray 
absorption column (\nh) and quasar redshift
(z), in the sense that intrinsic \nh\ increases with z. Furthermore in this
sample, the correlation does not depend on any calibration
effects nor the rest-frame of the absorbing column.}

\item{A decrease of iron K line equivalent width with
increasing radio-loudness. The
interpretation is that the strength of the reflection disk component (and
therefore any contribution from the iron K line) is diminished, due to
Doppler boosting of the X-ray continuum by the relativistic jet, in the
core-dominated radio-loud quasars.}

\end{itemize}

However we have seen several new effects and correlations in this paper,
that were not reported in our previous R97 sample:-

\begin{itemize}

\item{In the R97 paper, a correlation was found whereby the photon
index for the radio-quiet objects increased with X-ray luminosity. 
No such correlation was found in this paper, over a large range of both
redshift and luminosity, for the radio-quiet sources. The difference may be
due to the increase in sample size (from 9 radio-quiet quasars in R97 
to 27 in the present paper).}

\item{Two correlations were found involving the iron K emission line.
Firstly the strength of the iron K emission was observed to decrease with
luminosity (i.e. an `X-ray Baldwin' effect), regardless of whether the
objects are radio-loud or radio-quiet. In addition the energy (or
ionisation) of the iron line was found to increase with luminosity. Both of
these trends confirm the result found in the Nandra \et (1997b) paper,
whereby the composite line profiles for AGN tend to be weaker at higher
luminosities as well as narrower and shifted bluewards of 6.4 keV. The
`X-ray Baldwin effect' was first proposed for AGN on the basis of Ginga
data by Iwasawa \& Taniguichi (1993).}

\item{A new effect is found whereby the strength of the Compton reflection
`hump' is weaker in the most luminous quasars at high redshifts
(z$>$1). The effect is observed not only in the jet-dominated radio-loud
sources, but also in the {\it radio-quiet} quasars. This finding is consistent
with the `X-ray Baldwin effect' discussed above, and suggests that as a
whole the {\it neutral} disk reflection component in the high luminosity
quasars is generally weaker than in the lower luminosity sources such as
the Seyfert 1s.}

\item{A trend has also been found for the radio-quiet quasars in this
sample, whereby the X-ray (2-10 keV) 
photon index increases with decreasing optical
H$\beta$ width. Thus the quasars with the steepest X-ray spectra tend to
have the narrowest H$\beta$ FWHM. This trend has previously been found in
the lower luminosity Seyfert 1s (e.g. Brandt \et 1997), but has not been
reported before for the more luminous quasars.}

\item{Soft X-ray excesses are also found a significant
proportion (9) of the low z quasars in this sample. Interestingly the
majority of the quasars with strong soft excesses are those with the
narrowest optical H$\beta$ widths (where H$\beta$ FWHM $<2000$~km/s).}

\item{The temperatures of the soft X-ray excesses in this paper vary in the
range between kT = 100 - 300 eV, for simple blackbody fits. In a 
majority of the cases the
temperatures are probably too hot to result by direct thermal emission from
the putative quasar accretion disk. Instead, one possibility is that
the soft excess
originates via thermal Comptonisation of UV photons from the disk in a hot
corona. Another possibility is that the `soft excess' results
from reprocessing. In particular emission and/or reflection from the 
surface of a
highly ionised inner accretion disk could reproduce the observed excess in
soft X-ray flux.} 

\item{A systematic search has been carried out for the presence of warm or
ionised absorbers in this quasar sample. Only a smaller number were found.
We confirm the presence of a warm absorber in three previously reported
cases (PG 1114+445, IRAS 13349+243 and MR 2251-178). The only new warm
absorbers reported here are in the radio-quiet quasar PDS 456 and a
marginal detection in the radio-loud quasar PG 1425+267. Overall the
apparent rarity of warm absorbers in quasars may be due to  different
(higher) ionisation, a smaller covering fraction, or is perhaps just due to
the redshift effect.}

\end{itemize}

So how can we place all these observations facts into a general scheme for
quasars. Firstly the differences between radio-loud and radio-quiet quasars
seem relatively straightforward. In the radio-loud quasars, a strong
Doppler boosted emission component from the relativistic jet can account
for the higher luminosities, the generally flatter X-ray spectra as well as
the diminished iron K line and reflection component in these objects. One
question of real interest for future study is whether the 
central engine is the same in the radio-loud quasars as it is in the
radio-quiet quasars. For instance the structure of the accretion disk may
be different in the radio-loud quasars; sensitive studies of the reflection
component and iron line in the RLQs (i.e. with XMM) may help to
determine this. 

The properties of the {\it radio-quiet} quasars on the whole seem more complex.
As has been seen in this paper, there is little or no dependence on the
X-ray continua of quasars on luminosity and therefore presumably the black
hole mass. However perhaps the one driving factor responsible for the
individual properties of quasars may be the {\it fractional accretion
rate} $\dot{m}$ of the central engine (i.e. the ratio of mass
accretion to the Eddington rate - or the Eddington ratio). 
A high fractional accretion rate can result in the surface layers of the disk
becoming highly photoionised, which subsequently can have several effects
on the X-ray spectra. 
Depending on the degree of ionisation, ionised rather than neutral iron K
emission lines can dominate the disk reflection spectrum, as observed.
Furthermore at even higher ionisations, 
the neutral reflection component (and iron
line emission) can appear to be weaker, particularly if the disk is fully
ionised to several Thomson depths (e.g. Nayakshin \et 1999, Ross \et 1999),
also in agreement with the apparent properties of the more luminous
radio-quiet quasars. A further effect is that, for high Eddington ratios,
stronger soft X-ray emission can be produced. This may partly arise as the
intrinsic thermal emission from the disk can become stronger (Ross, Fabian
\& Mineshinge 1992). Another possibility is that as the disk is more highly
ionised, it can become more reflective at soft X-ray energies,
producing a steepening of the X-ray spectrum at low energies (i.e. as a
result of the ionised disk reflection component). Thus a high
accretion rate (relative to Eddington) may explain the
strong soft excesses in some
of the objects considered earlier.

It is also interesting to return to the question of the dichotomy between
the broad and narrow line quasars that was considered earlier. It has been
postulated in the literature that the narrow optical H$\beta$ lines may be
an indicator of a high accretion rate (Pounds \et 1995, Laor \et 1997). If
this is correct this could indeed account for the differences between the 2
types of objects. A high accretion rate may
explain the strong soft excesses observed both in the narrow-line
quasars (6 out of 8 objects in this sample) and also the lower luminosity
narrow-line Seyfert 1s (NLS1s) in other samples (Vaughan \et 1999, Leighly
1999). As explained this can be caused by increased intrinsic disk emission
or an increase in the reflectiveness of the disk in the soft X-ray band.
Also if the intrinsic disk emission is stronger, this can also account for
the steeper 2-10 X-ray slopes, via increased Compton cooling (Pounds \et
1995). Additionally some of the narrow-line quasars also show evidence for
ionised iron K emission, also indicative of a high ionisation disk and thus
a high accretion rate, 
although the evidence is tentative so far (also see
Vaughan \et 1999). So the narrow-line quasars (as
well as the NLS1s) may radiate at a relatively high fraction of the Eddington
rate, whereas in general the broad-lined quasars may be sub-Eddington,
perhaps similar to the normal Seyfert 1s, but with more massive central
black holes. An important question to ask is whether there are any
narrow-line quasars at higher redshifts (z$>1$).

Finally the amount of soft X-ray absorption towards
quasars was found to increase with redshift (also see Fiore \et 1998
for a similar analysis of \rosat\ quasars). 
This correlation is apparently robust, even when calibration effects
and uncertainties in the amount of local absorption are taken into
account. The main question that has arisen
from this, is whether this absorption is intrinsic to the quasars or
whether it originates from line-of-sight matter. Given the low-number
density of high column systems (such as damped Ly-$\alpha$ systems)
that could cause appreciable X-ray absorption (O'Flaherty \& Jakobsen
1997), the most likely scenario is that the bulk of this absorbing material
is local to the quasars or the host galaxy environment. As there seems to
be a comparative lack of
absorption in some radio-quiet quasars (see Fiore \et 1998), this
excess \nh\ may be associated with radio-loud quasars (also see Sambruna
\et 1999). 
However further data (with XMM and Chandra)
is required to determine the exact location and cause of this absorption.

\section*{ Acknowledgments }
We thank the \asca\ support teams, at GSFC and ISAS, for their
help. In particular, thanks to Ken Pounds and Simon Vaughan for proof 
reading the paper and providing useful discussions. We also
thank the anonymous referee for providing suggestions to improve the paper.
This research made use of data obtained from the Leicester Database
and Archive Service (LEDAS) at the Department
of Physics and Astronomy, Leicester University, UK, and the High Energy
Astrophysics Science Archive Research Center (HEASARC), provided by
NASA's Goddard Space Flight Center.

\label{lastpage}
\bsp

\end{document}